\documentclass[iop,apj]{emulateapj}
\usepackage{amsmath,amssymb,amstext}

\usepackage[breaklinks,colorlinks,citecolor=blue,linkcolor=magenta]{hyperref} 

\usepackage[all]{hypcap} 

\usepackage{natbib}
\usepackage[usenames, dvipsnames]{color}
\shorttitle{Electron Acceleration during the Eruption of a Flux Rope}
\shortauthors{E. Carley et al.}

\begin{document}

\title{ Radio Diagnostics of Electron Acceleration Sites During the Eruption \\ of a Flux Rope in the Solar Corona }

\author{Eoin P. Carley$^{1,2}$, Nicole Vilmer$^{2}$ and Peter T. Gallagher$^{1}$}
~
\affil{$^{1}$Astrophysics Research Group, School of Physics, Trinity College Dublin, Dublin 2, Ireland.\\
	\and
	$^{2}$LESIA, Observatoire de Paris, PSL Research University, CNRS, Sorbonne Universit\'{e}s, UPMC Univ. Paris 06, Univ. Paris Diderot, Sorbonne  Paris Cit\'{e}, 5 place Jules Janssen, 92195 Meudon, France. \\
              \email{eoin.carley@obspm.fr}}
              

\begin{abstract}

Electron acceleration in the solar corona is often associated with flares and the eruption of twisted magnetic structures known as flux ropes.
However, the locations and mechanisms of such particle acceleration during the flare and eruption are still subject to much investigation. Observing the exact sites of particle acceleration can help confirm how the flare and eruption are initiated and how they evolve.
Here we use the Atmospheric Imaging Assembly to analyse a flare and erupting flux rope on 2014-April-18, while observations from the Nan\c{c}ay Radio Astronomy Facility allows us to diagnose the sites of electron acceleration during the eruption.
Our analysis shows evidence for a pre-formed flux rope which slowly rises and becomes destabilised at the time of a C-class flare, plasma jet and the escape of {\color{black}$\gtrsim$75\,keV} electrons from rope center into the corona.
As the eruption proceeds, continued acceleration of {\color{black}electrons with energies of $\sim$5\,keV} occurs above the flux rope for a period over 5 minutes. %
At flare peak, one site of electron acceleration is located close to the flare site while another is driven by the erupting flux rope into the corona at speeds of up to 400\,km\,s$^{-1}$. Energetic electrons then fill the erupting volume, eventually allowing the flux rope legs to be clearly {\color{black} imaged from radio sources at 150--445\,MHz}. 
Following the analysis of \citet{joshi2015}, we conclude that the sites of energetic electrons are consistent with flux rope eruption via a tether-cutting or flux cancellation scenario inside a magnetic fan-spine structure. 
In total, our radio observations allow us to better understand the evolution of a flux rope eruption and its associated electron acceleration sites, from eruption initiation to propagation into the corona. 
\end{abstract}

\maketitle


\section{Introduction}

Flares and coronal mass ejections (CMEs) are thought to result from magnetic energy release in the solar corona, often involving the destabilisation of a twisted magnetic structure known as a flux rope \citep{chen2011, webb2012}. This activity may be accompanied by the acceleration of energetic particles \citep{lin2000a, kahler2007, lin2011}. However, there is ongoing debate on exactly where, when and how the particle acceleration occurs during flaring and eruption. Observing the locations of {\color{black}energetic electrons} during an eruptive event may help confirm how the electrons are accelerated, how the eruption proceeds, and also help in identifying which models of solar eruptive activity are correct.

The observation of the sites of electron acceleration during flaring or eruptive activity in the corona has traditionally been made using radio observations (see \citet{pick2008} for a review). 
Some of the longest known signatures of particle acceleration in the corona are type III radio bursts \citep{wild1959}, now believed to be from energetic electrons causing plasma emission as they {\color{black}propagate through} the corona \citep{paesold2001, yan2006, chen2013}. 
Type IIIs are amongst various types of radio bursts generated by energetic electrons that are accelerated during both flares and small scale eruptions in the form of plasma jets \citep{aurass1994, kundu1995, nitta2006, klassen2012, chen2013a}. Sites of energetic electrons (radio sources) are also known to be closely associated with larger scale eruptions such as plasmoids and sigmoids \citep{kundu2001, khan2002, marque2002}. 
Such eruptive activity often shows the {\color{black}sites of energetic electrons} to be located close to the underlying active region or moving with the {\color{black}erupting structure} itself \citep{pick2005, bain2014}, originally observed as flare continua and moving type IV bursts \citep{robinson1975, pick1986}.
Electron acceleration sites may also be located on the boundaries of CMEs, located at its nose or flanks. \citep{zimovets2012, bain2012, carley2013,  zucca2014, salas2016}.
In the later stages of an eruptive event, {\color{black}the radio emission from energised electrons} can be observed to be within the erupting structure, allowing sources of plasma emission at the CME legs to be imaged \citep{maia1999, huang2011}. 
In very rare cases, the energised electrons fill the entire volume of the CME and interact with the magnetic field to produce gyrosynchrotron emission in the metric/decimetric domain, allowing observation of what is generally known as a `radio CME' \citep{bastian2001, maia2007, demoulin2012}.

Theoretical models of solar eruptive activity often include a variety of sites of magnetic reconnection and shocks \citep{chen2011}, implying a variety of possible particle acceleration sites during eruption. The models may be unique in where they predict these sites to occur in relation to the erupting flux rope. For example, the magnetic breakout model specifically predicts a site of reconnection above a flux rope \citep{antio99, lynch2004}, while the tether-cutting or flux cancellation models may both build and release the flux rope via reconnection quite close to the rope center \citep{vanboo1989, moore2001}. During the final stages of propagation of the erupting structure, nearly all models predict the development of a current sheet below the main body of the eruption. This is generally known as the `standard' or CSHKP model, and predicts reconnection, shocks and particle acceleration in this current sheet \citep{carmichael1964, sturrock1966, hirayama1974, kopp1976}. Elsewhere during the eruption, models also predict particle acceleration from interchange reconnection or shocks driven at the outer boundaries of the erupting structure as it propagates into the corona \citep{kozarev2011, schmidt2012, masson2013}.

While theoretical models predict a variety of possible electron accelerations sites during flux rope eruption, the radio observations provide a means for detecting theses sites. 
However, observing the sites of electron acceleration simultaneously with flux rope observations has proven difficult in the past. It is only recently that the Atmospheric Imaging Assembly \citep[AIA;][]{lemen2012} has made available high spatial and temporal resolution observations of flux rope signatures in the corona e.g., twisted sigmoids with temperatures of $\gtrsim$10\,MK \citep{zhang2012, cheng2015, joshi2015, song2015}. Hence, there is now the possibility to combine the high time and spatial resolution flux rope observations with radio imaging and dynamic spectra to explore where, when and how the electron acceleration occurs during the eruption of such a body and to compare this to what theoretical models predict. 

In this paper, we examine an eruptive event from 2014-April-18. This event has previously been studied by \citet{joshi2015} and \cite{cheng2015} using X-ray and EUV imaging and UV spectroscopy, respectively. They identify a flux rope and multiple hypothesized sites of reconnection (potentially associated with electron acceleration) during eruption. 
Here we attempt to identify these sites of electron acceleration using radio imaging from the Nan\c{c}ay Radioheliograph \citep[NRH;][]{kerdraon1997} combined with metric and decimetric radio spectrography. We show strong observational evidence of {\color{black} energized electrons produced from an eruptive mechanism which closely resembles a tether-cutting or flux cancelation model. This is followed by electron acceleration from reconnection above the erupting structure in the surrounding magnetic environment.} Sites of electron acceleration are then both associated with the flare and driven by the {\color{black}erupting structure} itself. We also show how electrons then fill the erupting volume, making visible at radio wavelengths the CME legs. In total, we reveal in unprecedented detail the sites and kinds of electron acceleration occurring throughout this event, from eruption initiation to propagation into the corona.

\begin{figure}[!t]
    \begin{center}
    \includegraphics[scale=0.35, trim=2.5cm 0.8cm 2cm 1.0cm]{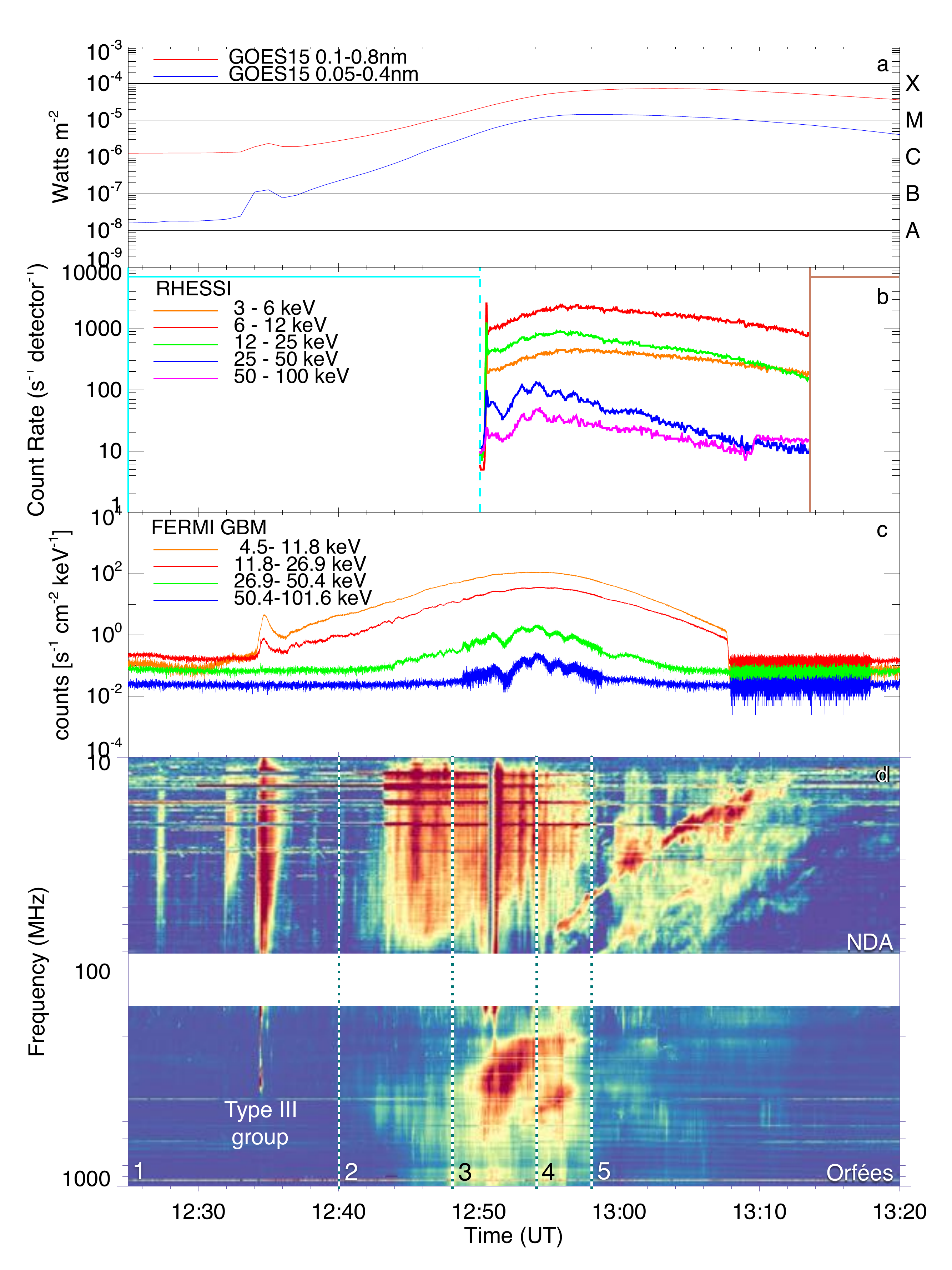}
    \caption{ (a) GOES X-ray light curves showing the initial C-class flare at $\sim$12:35\,UT. This is followed by an M7.3 class flare peaking at $\sim$13:00\,UT. (b) RHESSI X-ray flux observations from 3-100\,keV. Data gaps due to RHESSI night and the South Atlantic Anomaly are indicated by blue and brown lines, respectively. (c) FERMI GBM light curves showing emission from 4.5-101.6\,keV. (d) Radio dynamic spectra from NDA and Orf\'{e}es covering 10--1000\,MHz. 
    Based on the observations below, the dynamic spectrum is split into five periods, indicated by the vertical dashed lines.}
    \label{fig:xray_radio}
    \end{center}
\end{figure}
\section{Observations}

\begin{figure}[!ht]
    \begin{center}
    \includegraphics[scale=0.385, trim=1cm 0cm 0cm 0cm]{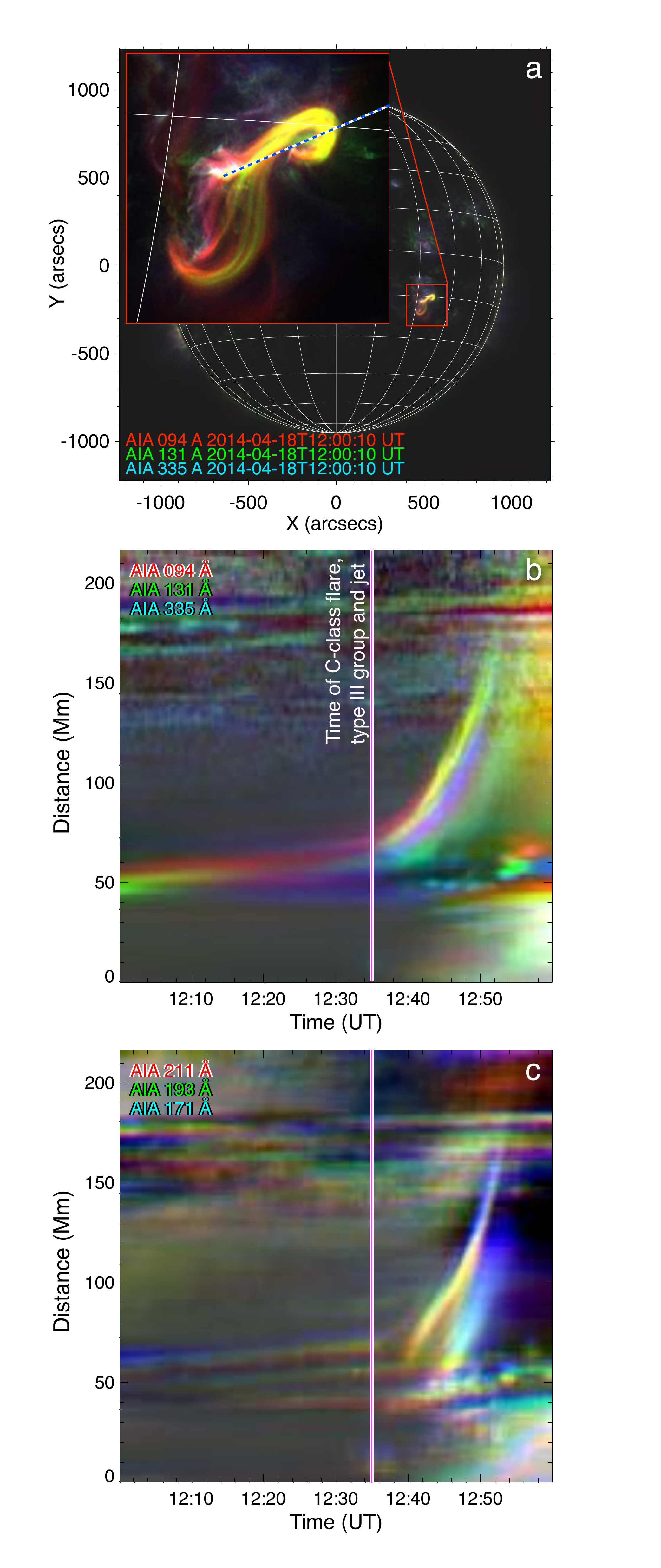}
    \caption{ (a) Three colour image of full sun using AIA 94, 131, 335\,\AA. The inset shows the twisted flux rope structure with emission dominating in the 94 and 131\,\AA~channels ($\sim$6 and 10\,MK, respectively). This structure exists for several hours prior to eruption. A distance-time map is constructed along the dashed blue-white line. (b) Distance-time map of the AIA hot channels {\color{black}(see text)}. The initial slow motion can be seen as the slow rise of the green-red feature from 50\,Mm. The initial C-class flare and metric type III group occur at 12:34:30\,UT, marked by vertical white-pink line. (c) Three-color map of the cooler AIA channels showing acceleration phase at $\sim$12:40\,UT.}
    \label{fig:aia_dt_maps}
    \end{center}
\end{figure}

In the following we present radio dynamic spectra from the Nan\c{c}ay Decametric Array \citep[NDA;][]{lecacheux2000} between 10-80\,MHz and Nan\c{c}ay Radio Observatory's 
Orf\'{e}es\footnote{\emph{Orf\'{e}es: Observation Radiospectrographique pour FEDOME et l'\'{E}tude des \'{E}ruptions Solaires}, where FEDOME is the \emph{F\'{e}d\'{e}ration des Donn\'{e}es M\'{e}teorologiques de l'Espace.}} spectrograph (a Fast Fourier Transform spectrometer providing dynamic spectra between 140--1000\,MHz at 0.1 second nominal time resolution). To produce radio images, all NRH data were CLEANed and integrated to 1 second time resolution using the standard NRH packages available in \emph{SolarSoft}, providing observations at 150, 173, 228, 270, 298, 327, 408, 432 and 445\,MHz. Figure~\ref{fig:xray_radio} summarizes the X-ray and radio dynamic spectra observations of the event from 12:25--13:20\,UT including (a) soft X-ray light curves from the \emph{GOES} spacecraft, (b) 3-100\,keV X-ray light curves from the \emph{Reuven Ramaty High-Energy Solar Spectroscopic Imager} \citep[RHESSI;][]{lin2002}, (c) 4.5--101.6\,keV X-ray light curves from the \emph{FERMI} Gamma Ray Burst Monitor \citep[GBM;][]{meegan2009} and (d) dynamic spectra from NDA and Orf\'{e}es. We have separated the observations into periods 1-5, marked by the vertical dashed lines on the dynamic spectra. We describe these periods separately in the following sections.

\subsection{Electron beam generation during eruption initiation}

The 2014-April-18 event began with a small C-class flare at $\sim$12:35\,UT, followed by an M7.3 flare observed by the GOES spacecraft, Figure~\ref{fig:xray_radio}(a). In radio dynamic spectra in `period 1', two decametric type IIIs are observed in NDA, followed by a metric `type III group' at 12:34:30\,UT which {\color{black} starts in the decimetric domain (frequencies above $\sim$300\,MHz)}, as observed in Orf\'{e}es and indicated on Figure~\ref{fig:xray_radio}(d). {\color{black}The type IIIs in this group occur at the time of the initial C-class flare.}

During this time, a twisted loop structure located at NOAA active region 12036 was observed in AIA 94, 131, 335\,\AA~channels (peak temperature responses at $6.3\times10^6$\,K, $1.6\times10^7$\,K and $2.5\times10^6$\,K, respectively, and hereafter called `hot channels'), see Figure~\ref{fig:aia_dt_maps}(a). This is a three-color map in which three AIA passbands are represented by the red, green and blue (RGB) channels of a color image. Purely white areas of such a map indicate equal intensity contribution from all AIA passbands, while a primary color (or combination of primaries) indicates one (or two) channels being dominant in intensity. At 12:00\,UT we observed the twisted structure is composed of yellow (94, 131\,\AA) and red (94\,\AA) strands, indicating temperatures in the range of 6--10\,MK. 
The studies of \citet{joshi2015} and \cite{cheng2015} provide evidence that this structure is a flux rope, showing its development hours before eruption.  In our study, we concentrate on 12:00\,UT onwards, when the flux rope has already been formed and poised for eruption. 

In order to determine any pre-eruptive motion of this structure we constructed a distance-time (d-t) map by extracting intensity traces along a straight line in the north-western part of the rope, shown in Figure~\ref{fig:aia_dt_maps}(a) inset. We did this for both the AIA hot channels and the 171, 193 and 211\,\AA~channels (these channels have peak temperature responses at $6.3\times10^5$\,K, $1.2\times10^6$\,K and $2\times10^6$\,K, respectively, and are called `cooler channels' hereafter). Figure~\ref{fig:aia_dt_maps}(b) shows the hot d-t map; the flux rope is present from 12:00\,UT, seen as the green and red emission rising slowly from 50\,Mm. At 12:35\,UT acceleration begins, 
coincident with the initial C-class flare and type III group radio burst, marked by the white-pink vertical line on the distance time maps. The cooler channels, shown in Figure~\ref{fig:aia_dt_maps}(c), show that the erupting loops only become bright at $\sim$12:40\,UT, possibly due to the structure cooling to these temperatures $\sim$5 minutes into the acceleration phase. The acceleration is visible until $\sim$12:50\,UT when the structure reaches a velocity of $>$200\,km\,s$^{-1}$ and fades from all channels (a detailed kinematical analysis of the event is given in the Kinematics Section~\ref{section:flux_rope_kins} of this paper).

The eruption may be examined more closely in movie\_1.mpg. At $\sim$12:31\,UT we observe a pinched `Y'-shaped structure at rope center, followed by the brightening from the flare at $\sim$12:35\,UT, then the expulsion of material in the form of a jet with a velocity of 120\,km\,s$^{-1}$ \citep{joshi2015}. The flare brightening and jet take place at the time of the type III group in the dynamic spectrum.

In order to understand the relationship between the signatures of electron acceleration from the radio dynamic spectra (type III group) and the erupting jet and flux-rope as seen in AIA, we use images provided by NRH in all frequency bands. Figure~\ref{fig:fluxrope_typeIII}(a) shows a three-color combination of AIA hot channels clearly showing the flux rope structure and jet. Figure~\ref{fig:fluxrope_typeIII}(b) shows the same image with a wider field of view and over plotted with radio contours from NRH. 
The image is at the same time as the type III group observed in dynamic spectra. In movie\_2.mpg, throughout this type III group, the radio contours are observed to align south of the rope on three separate occasions (three separate type III bursts), albeit in slightly different directions. This shows that beams of electrons were accelerated close to the flux rope along with the plasma jet. 
 A frequency-time trace of the type III group from the NDA and Orf\'{e}es dynamic spectra gives an electron energy of {\color{black}$\gtrsim$75\,keV} using the electron density model of \citet{sthilaire2013}. The details of this calculation are given in Sections~\ref{discussion:fan_spine}.


\subsection{Electron beam generation during eruption acceleration}

\begin{figure}[!t]
    \begin{center}
    \includegraphics[scale=0.54, trim=3cm 0.5cm 3cm 0.5cm]{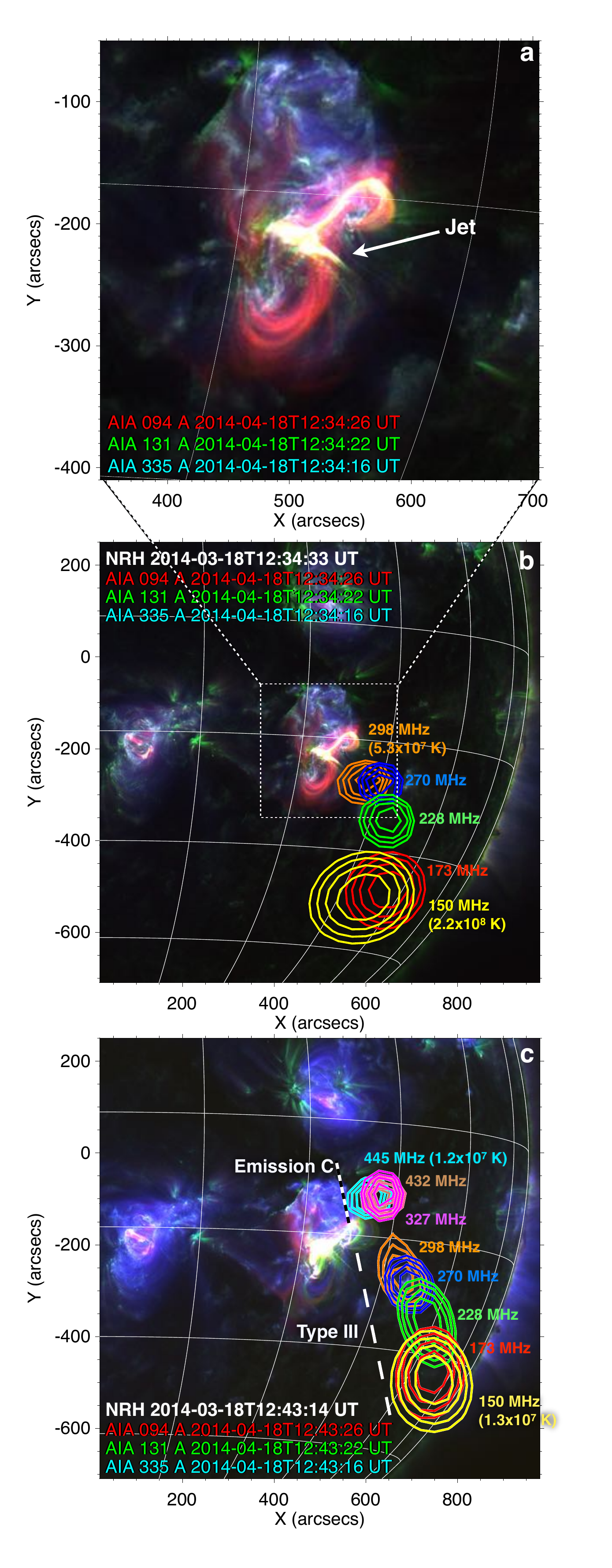}
    \caption{ (a) AIA three colour image of 94, 131, 335\,\AA~channels showing a highly twisted structure with a jet (indicated by the arrow) occurring at its center. (b) Same image as (a) but with a wider field of view and NRH contours from 327--150\,MHz over plotted and maximum and minimum brightness temperature indicated on the corresponding frequency. The sources occur at the same time as the type III group at 12:34:30\,UT and make a roughly straight line above the jet with higher frequencies at lower altitude. (c) The radio sources occur above the north-west loop of the rope. This activity last from 12:43:30--12:46:00\,UT (see movie\_2.mpg). The dashed lines indicate the frequencies of type III emission and Emission `C' shown during period 2 of the dynamic spectrum in Figure~\ref{fig:source_motion_imgs}.}
    \label{fig:fluxrope_typeIII}
    \end{center}
\end{figure}

\begin{figure*}[!t]
    \begin{center}
    \includegraphics[scale=0.5,  trim=3cm 0cm 0cm 0cm]{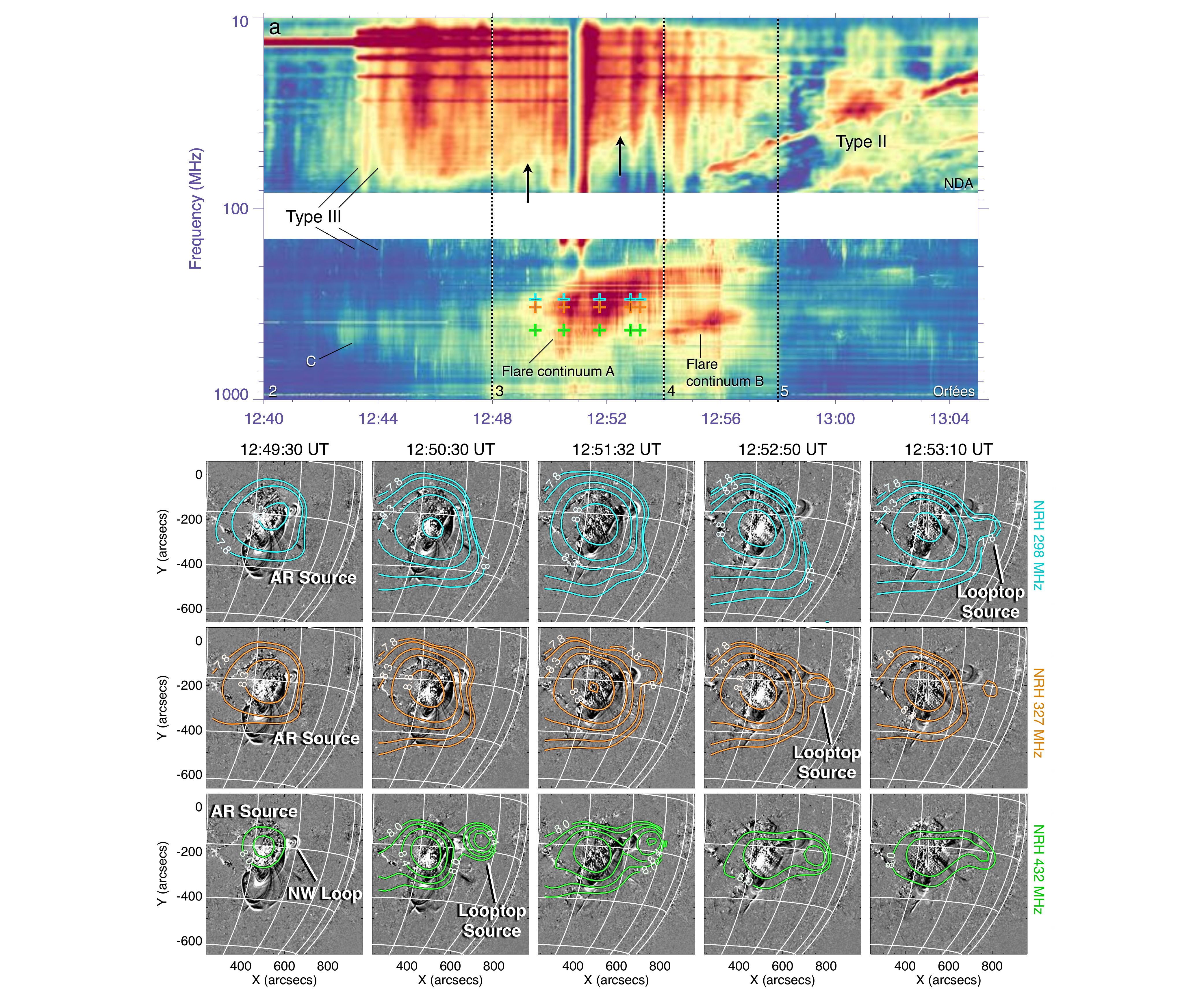}
    \caption{(a) Orf\'{e}es and NDA dynamic spectrum observations from $\sim$1000 to 10\,MHz during periods 2, 3, 4 and 5 as indicated by the vertical dashed lines. Emission `C' and type IIIs are observed in period 2 (imaged in Figure~\ref{fig:fluxrope_typeIII}(c)). In period 3 `Flare Continuum A' begins, while the starting frequencies of the type IIIs observed in NDA show a slow drift towards lower frequencies, indicated by the black arrows. The coloured crosses highlighted on the dynamic spectrum correspond to the frequencies and times of the NRH contours below the dynamic spectrum, overlaid on AIA 171\,\AA~running ratio images (NRH contours are in $log_{10}(T_B $\,[K])). At each frequency, two sources are identified in these images; a stationary (AR) source located above the active region and a much weaker looptop (LT) source located above the north-west loop (`NW Loop', as indicated) of the rope. During period 4 a separate `Flare Continuum B' can be observed in Orf\'{e}es around 400\,MHz.}
    \label{fig:source_motion_imgs}
    \end{center}
\end{figure*}

Following this type III group, a second set of type IIIs is observed to occur in NDA starting from $\sim$12:43\,UT onwards -- during `period 2' in the dynamic spectrum in Figure~\ref{fig:source_motion_imgs}(a). Although many of these type IIIs begin in the NDA range below 80\,MHz, some can be observed in Orf\'{e}es to start at frequencies as high as {\color{black}$\sim$200\,MHz}, as indicated on the dynamic spectrum\footnote{\color{black}Orf\'{e}es uses a single dish receiving system and is therefore far less sensitive than the NRH or NDA. The type IIIs in Orf\'{e}es are therefore weaker in intensity than those observed in NDA and appear to start from a smaller frequency in Orf\'{e}es (200\,MHz) than in NRH (298\,MHz).}. Also observed during period 2 in Orf\'{e}es is an emission composed of continua and pulsations between $\sim$300-500\,MHz, labelled emission `C'. At this time, the 327, 432, 445\,MHz radio sources are clustered above the north-west loop of the rope, while the sources at 298\,MHz and below show a roughly southerly alignment, shown in Figure~\ref{fig:fluxrope_typeIII}(c). The radio sources show this configuration from $\sim$12:40:30\,UT until $\sim$12:46:00\,UT (movie\_2.mpg). This shows that the source of emission C and the type IIIs observed in Orf\'{e}es during period 2 have a close spatial relationship. It is likely that emission C is produced by some {\color{black}electron acceleration} site at the north-west point of the rope, while the type IIIs may represent an escape of these electrons on open field lines, much the same as the mechanism outlined in \citet{paesold2001}. The frequency drift of these type IIIs was used to estimate electron beam speeds of $\sim$0.14\,c ($\sim$5\,keV) using the St. Hilaire model stated above. This may be interpreted as electron beams that escape into the corona due to the flux rope interaction with the surrounding magnetic environment at its north-west vicinity, which we discuss further in Section~\ref{discussion:fan_spine}.

\subsection{Energetic electrons during HXR flare peak}
\subsubsection{Flare Continuum A}

\label{obs:flare_continuum}
\begin{figure}[!t]
   \begin{center}
   \includegraphics[scale=0.40,  trim=2cm 0cm 0cm 0cm]{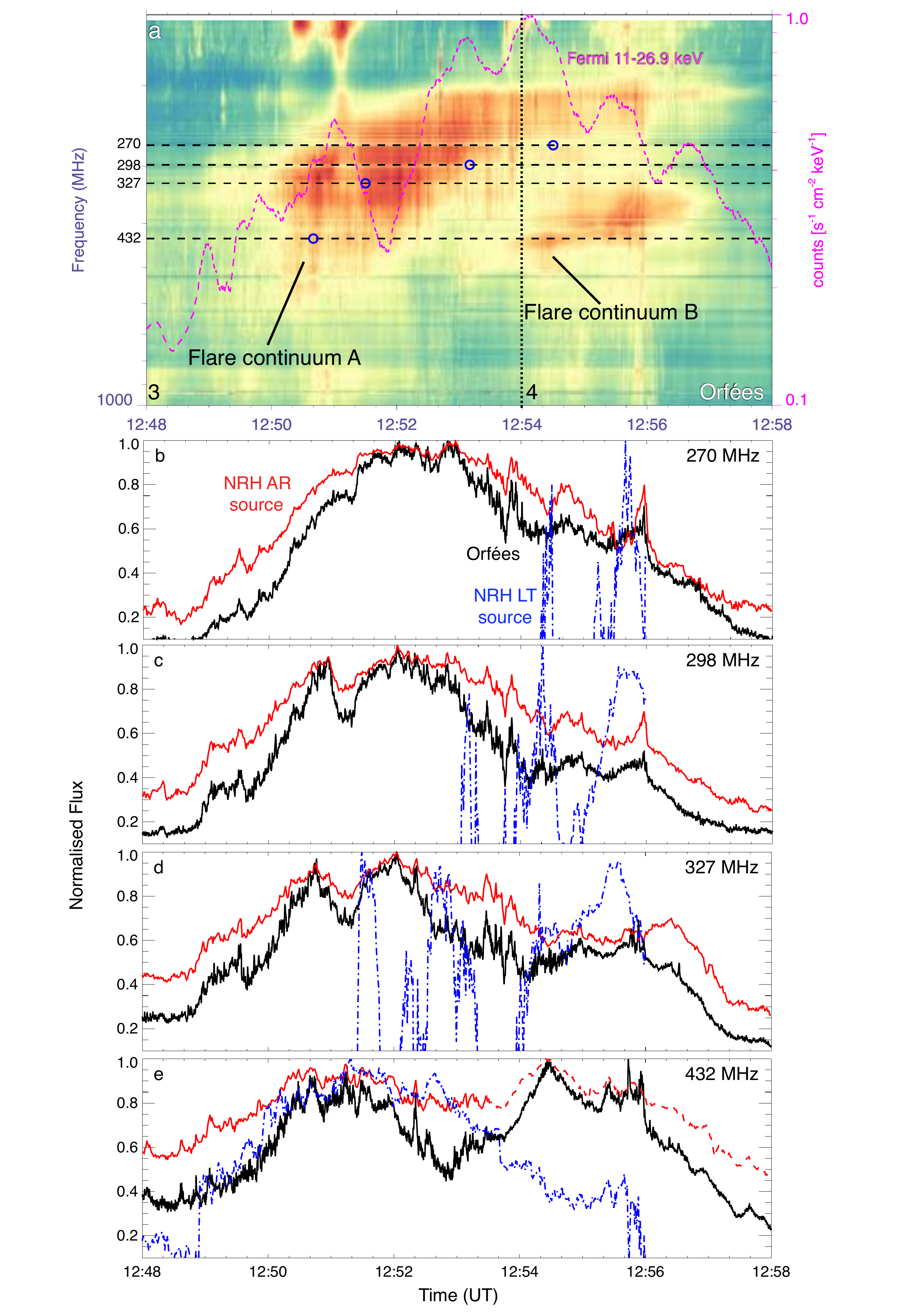}
    \caption{(a) Detail of the Flare Continuum A and B radio bursts in the Orf\'{e}es spectrogram. The dashed lines indicate the frequencies at which the light curves below are taken. The blue circles indicate the times of the first flux peak of the Looptop (LT) source from NRH images. The pink curve is FERMI GBM 11--26.9\,keV X-ray flux. (b)-(d) Comparison of normalized Orf\'{e}es flux with the normalized flux of the active region (AR) source and LT source as imaged by NRH - see Table~\ref{tab:table0} for flux normalization values. Comparing black and red curves, the stationary AR source is responsible for the majority of Flare Continuum A, while the LT source (blue dashed curves) is much more sporadic. (e) Both AR source and LT source are comparable to the Orf\'{e}es flux variation over time until 12:53\,UT, suggesting they were both involved in the flare continuum at 432\,MHz. Note the red curve changes from solid to dashed at $\sim$12:53\,UT, indicating that the AR source at this frequency diminishes and a new source appears (seen in Figure~\ref{fig:looptop}). }
    \label{fig:flux_comparison}
    \end{center}
\end{figure}

\begin{figure}[!t]
   \begin{center}
   \includegraphics[scale=0.34,  trim=0cm 2cm 0cm 3.0cm]{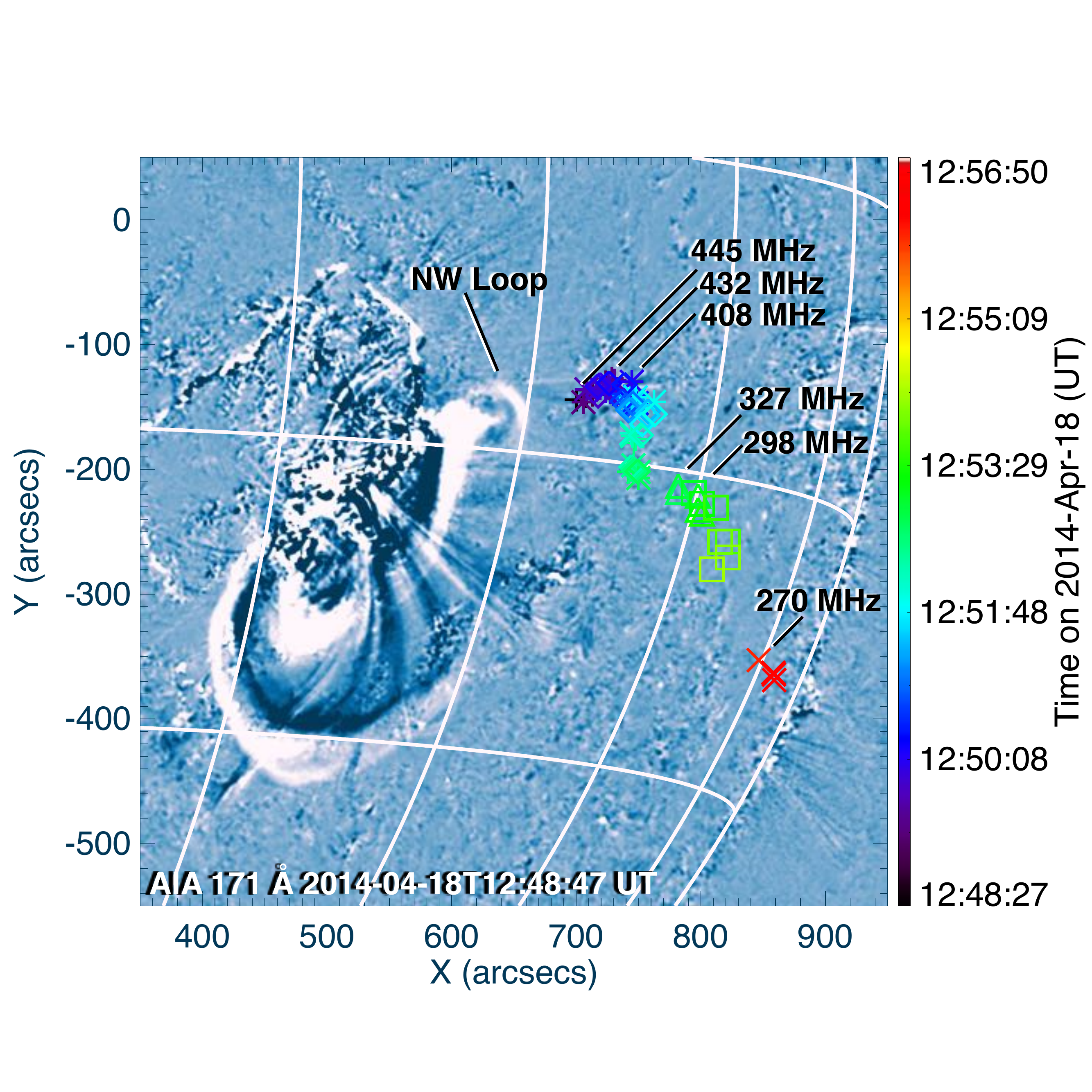}
    \caption{{\color{black}Motion throughout time (indicated by color) of the Looptop (LT) source observed at multiple NRH frequencies, over-plotted on an AIA 171\,\AA~running difference image at a single time to illustrate the position of these sources with respect to the north west (NW) loop. The first position of the source at each frequency is indicated. The sources have a speed of $\sim$395\,km\,s$^{-1}$ in the plane of sky. Although the sources occur at the time of  Flare Continuum A, they were too weak to be directly responsible for this radio burst.}}
    \label{fig:source_motion}
    \end{center}
\end{figure}

We next turn our attention to the radio burst labelled `Flare Continuum A' observed to start during period 3 in the Orf\'{e}es dynamic spectrum, shown in Figure~\ref{fig:source_motion_imgs}(a). The figure shows the burst is broadband, initially covering $\sim$300--600\,MHz, but with this frequency band slowly drifting toward lower frequencies over time. The coloured crosses indicate the times and frequencies of the NRH contours overlaid on the AIA 171\,\AA~images shown below the spectrogram. We observe that throughout the lifetime of Flare Continuum A a stationary source is observed at 228--445\,MHz above the active region (only three frequencies are shown here for simplicity), labelled `active region' or `AR' source in the left-most column of images. The AR source appears to be largely responsible for Flare Continuum A observed in Orf\'{e}es. However, there is also a second, smaller source which is observed above the north-west loop (`NW Loop') of the flux rope, labelled `Looptop' or `LT' source e.g., starting at NRH 432\,MHz at 12:50:30\,UT and appearing successively later at lower frequencies, as indicated in the images with the 327 and 298\,MHz contours. 

\begin{figure}[!t]
   \begin{center}
   \includegraphics[scale=0.37,  trim=0cm 0cm 0cm 0cm]{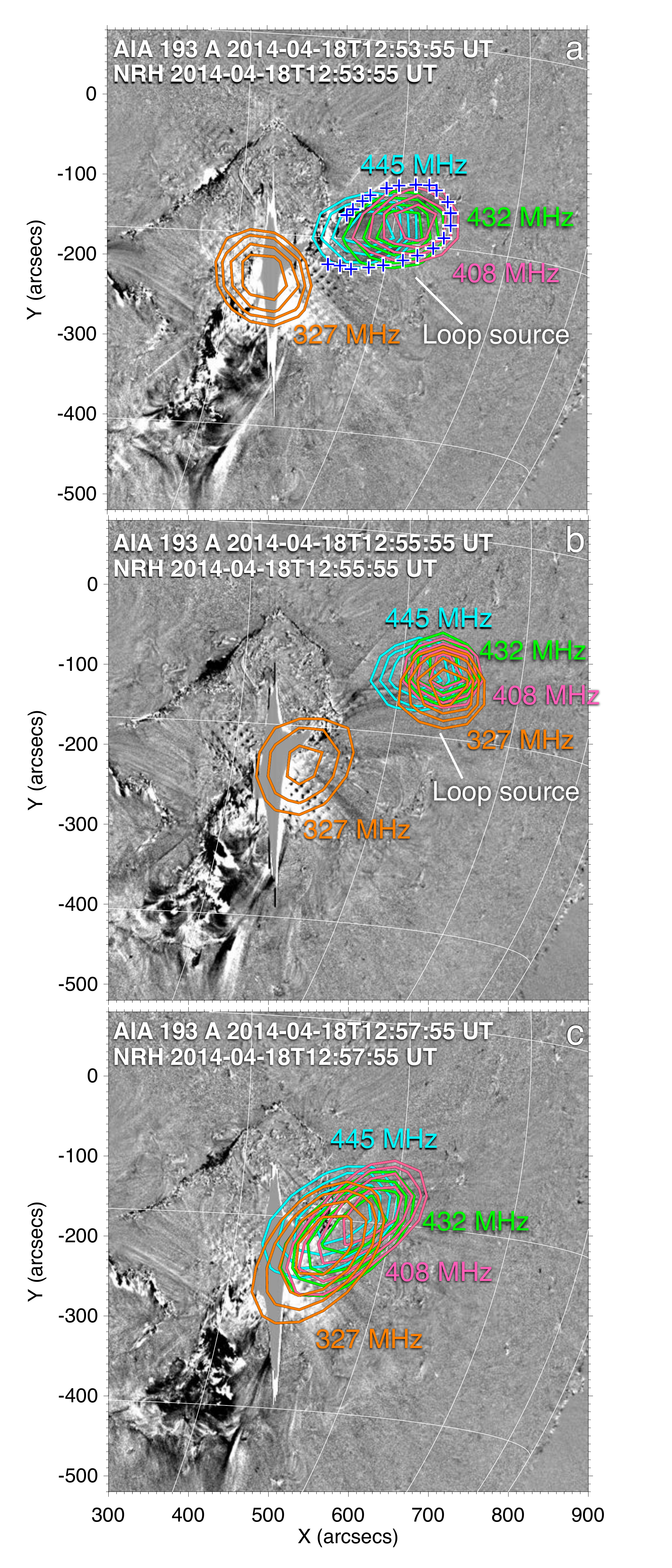}
    \caption{AIA 193\,\AA~running ratio images over plotted with NRH 327, 408, 432 and 445\,MHz. (a) The radio contours at 408\,MHz and above are now located inside the north-west loop of the rope (this loop is visible in the AIA image but obscured by the radio contours, hence it is demarcated by the blue-white crosses). The remainder of the rope is now indistinguishable in the images. The 327\,MHz source (and lower frequencies) is located above the active region. (b) The 408--445\,MHz sources then moves in the direction of the loop as it propagates westward. The 327\,MHz contours shows a source above the active region and one now clustered with the higher frequencies at the loop location. (c) The sources then begin to shift back towards the active region and remain there. Panels (a) to (b) show the radio sources resulting in the `Flare Continuum B' feature of Figure~\ref{fig:flux_comparison}(a).}
    \label{fig:looptop}
    \end{center}
\end{figure}

\begin{center}
\begin{table} [b!]
\centering   
\caption{NRH source flux densities}
    \begin{tabular}{lcc}
        \hline
        	      Frequency &   Active region source	  &  Looptop Source   \\
	        	(MHz)			 &   (SFU)	  &  (SFU)   \\
	  \hline   
			270 	&  172  	&  5  \\
			298  	&  255 	  	&  24  \\
			327	&  294	  	&  23	\\
			432  	&   219	  	&  141 	\\
			  \hline
	    \end{tabular}
	    \tablecomments{Maximum flux density of the AR and LT sources during their lifetime. The maxima occur at different times for each source and frequency. 1\,SFU $=10^4$\,Jansky$=10^{-22}$\,W\,m$^{-2}$\,Hz$^{-1}$}
        \label{tab:table0}
   
\end{table}
\end{center}
To determine how the AR and LT sources contribute to Flare Continuum A, we compare the flux density of these sources observed with NRH to the flux density observed with Orf\'{e}es at the same frequencies; the results are shown in Figure~\ref{fig:flux_comparison}. Panel (a) provides a zoom of the radio burst and (b)--(e) provide normalised flux comparisons at 270, 298, 327 and 432\,MHz . There is a much stronger relationship between Flare Continuum A flux (black curve) and the AR source flux (red curve). The LT source (blue curve) is sporadic and only appears after Flare Continuum A has begun. For example, we have marked with circles on the dynamic spectrum in Figure~\ref{fig:flux_comparison} the time at which the LT source reaches its first flux peak at each frequency. At frequencies below 327\,MHz the LT source only appears toward the end of Flare Continuum A. Furthermore, as shown in Table~\ref{tab:table0}, the LT source was much weaker in flux density than the AR source at most frequencies. It is only at 432\,MHz that the two sources are comparable in their maximum fluxes. Hence Flare Continuum A was primarily produced by the AR source, with some contribution by the LT source only at frequencies of $\sim$432\,MHz and above.

Previous observations of flare continua {\color{black}during flare peak do not always show} a frequency drift in dynamic spectra\footnote{{\color{black}As outlined in \citet{mclean1985} and \citet{pick1986}, there are different subclassifications of flare continua. Due to its occurrence at flare rise and peak, we identify this flare continuum as a flare continuum-M (FCM), also know as a flare continuum early (FCE). FCM (or FCE) do not usually show a drift in dynamic spectra. A separate type of continuum, known as FCII or type IVmB, often do show a drift, but this usually occurs slowly over a period of $\sim$1\,hour after the flare peak.}}.
However, interestingly, Flare Continuum A shows a drift to lower frequencies over time. This may be due to the active region source losing density over time and therefore emitting at increasingly lower plasma frequencies. 
Indeed, because of its drift, this flare continuum resembles previous observations of `drifting pulsating structures' (DPS) \citep{karlicky1994, khan2002, karlicky2005} or `type II precursors' \citep{klassen2003}. Those studies showed such features to be associated with moving radio sources in the corona. However, in the event studied here, the predominant radio source shows no motion. The drifting flare continuum is from a stationary site of electron acceleration associated with the flaring active region. It is therefore important to stress that frequency drift in dynamic spectra does not necessarily imply motion of the burst driver. Both radio dynamic spectra and radio imaging are needed to properly identify the origin of such drifting radio bursts.

Finally, we note that the LT source shows a consistent motion in the plane-of-sky (POS). Tracking this source through time and frequency in the NRH images, we see that it appeared at successively lower frequencies over time, indicating a driver with a velocity of $\sim$395\,km\,s$^{-1}$, as shown in Figure~\ref{fig:source_motion}. {\color{black}This source was likely driven by the expanding north-west loop of the rope as it erupts, similar in observation to \citet{klassen1999}, for example. The interaction of this expanding loop with its surrounding environment as it erupts would drive plasma emission at larger heights (lower frequencies) over time} - in Section~\ref{section:flux_rope_kins} we show this radio source kinematics and the flux rope kinematics are directly comparable. We emphasise that despite this radio source occurring at increasingly lower frequencies over time, it did not contribute directly to the frequency drift of Flare Continuum A since its flux is negligible, see Table~\ref{tab:table0}.

\begin{figure*}[!t]
    \begin{center}
    \includegraphics[scale=0.5, trim=0cm 0cm 0cm 0cm]{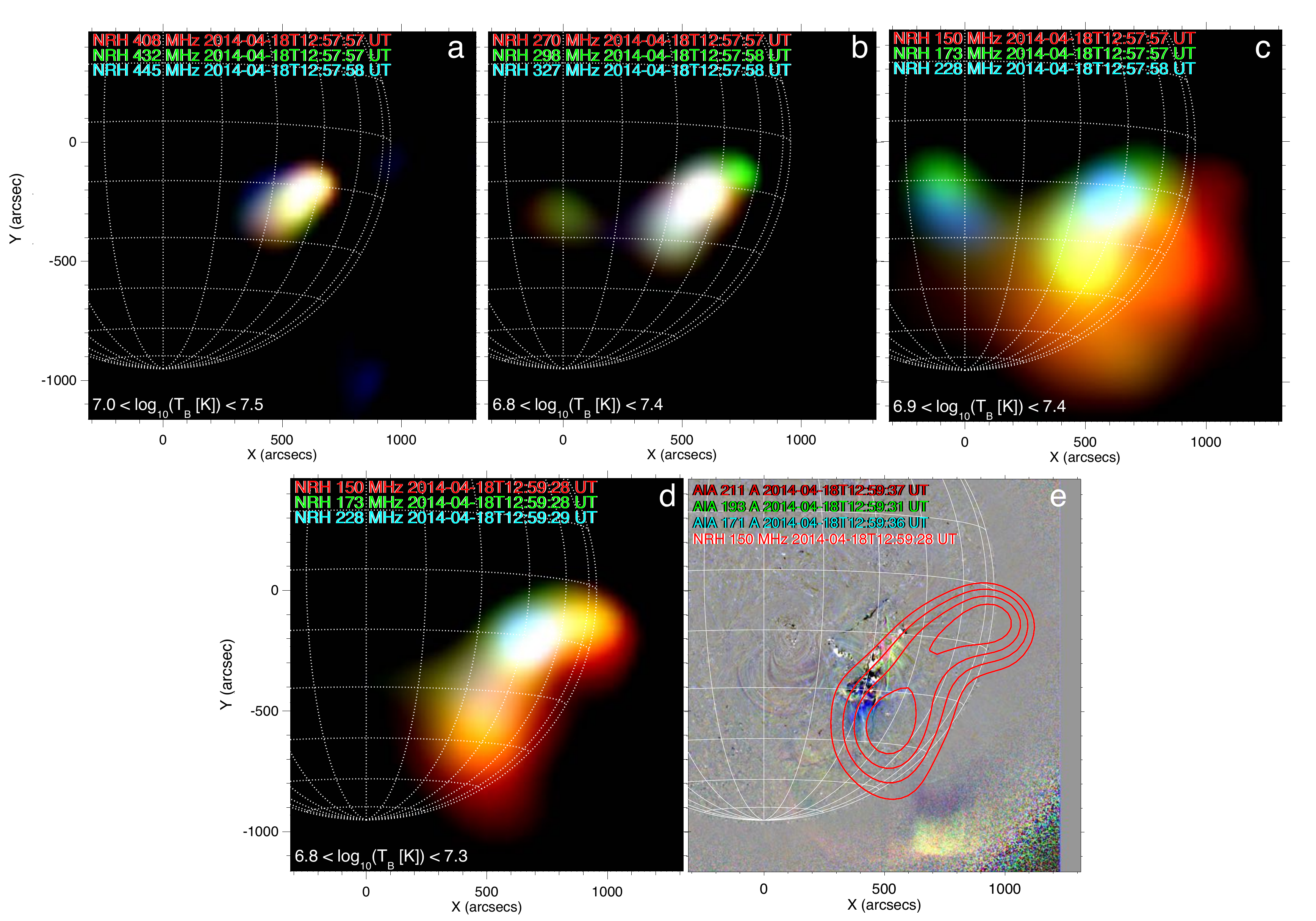}
    \caption{ NRH observations with frequencies displayed in RGB triplets. (a) 445, 432 and 408\,MHz, (b) 327, 298 and 270\,MHz, (c) 228, 173, 150\,MHz each at 12:57:57\,UT. In these images, any spatially coherent levels in brightness temperature are displayed as white i.e., equal amounts of red, green and blue. An increase in any one (or two) of the three images in the triplet is displayed as a primary colour (or combination of primaries). The brightness temperature scaling is indicated on each triplet. In panels (a) to (c) all radio emission is generally situated above the active region, with the emission at the lowest frequencies grouped in a `radio bubble'. (d) The lowest frequencies then develop into a `radio arc', while (e) {\color{black}displays an EUV front near the edge of the AIA field of view}, with NRH 150\,MHz contour for comparison.}
    \label{fig:nrh_aia_3col}
    \end{center}
\end{figure*}

\begin{figure}[!t]
    \begin{center}
    \includegraphics[scale=0.37, trim=3.0cm 0.5cm 3cm 0cm]{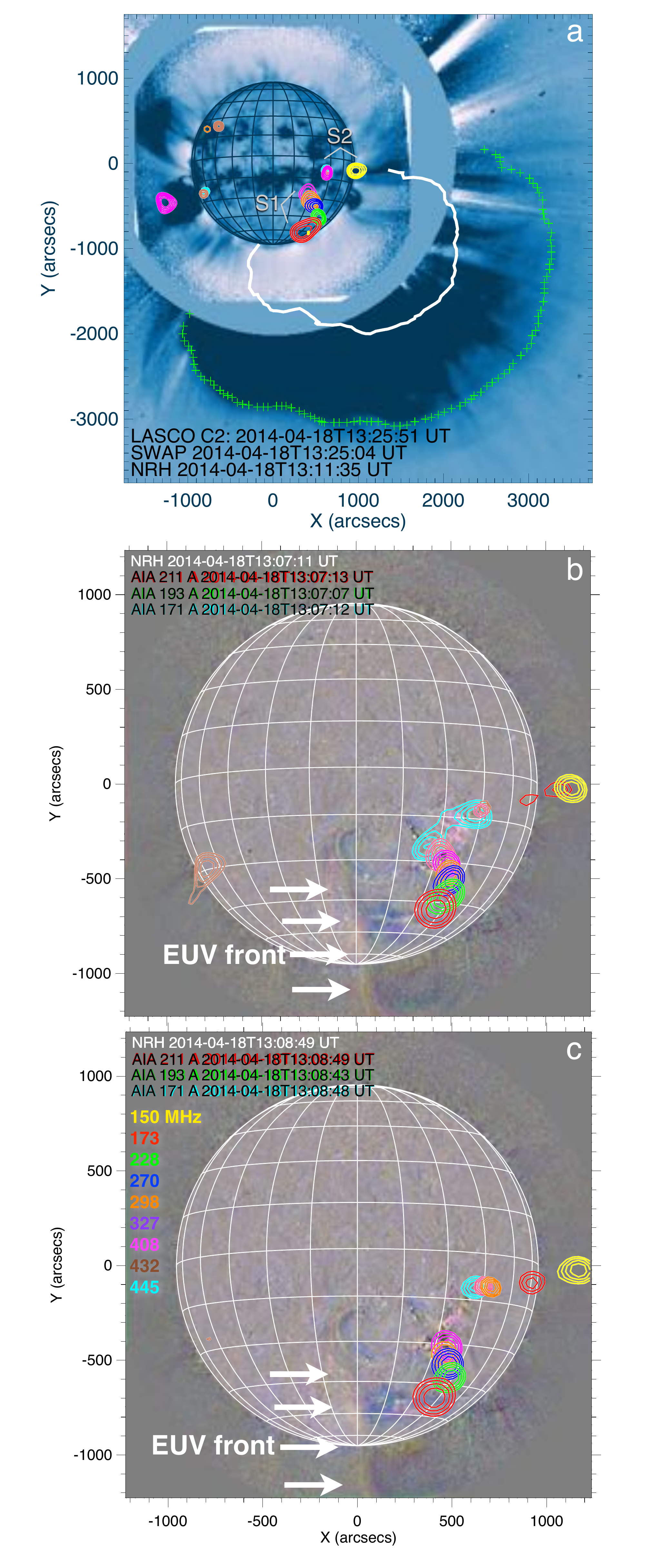}
    \caption{ (a) Combination of SWAP 174\,\AA~, NRH contours (colors/frequencies marked in panel (c)), and LASCO C2 image of the CME. The NRH contours are grouped into two segments `S1' and `S2'. The green crosses are the positions of the CME front chosen by point-and-click. The white line represents the back-extrapolation of this front using a speed of 1100\,km\,s$^{-1}$ and a propagation time of 14 minutes (time difference between NRH image and LASCO C2 image). The back-extrapolated front and NRH contours show a good spatial correspondence. (b) and (c) show the position of the radio contours (possible CME legs) and the EUV front propagating along the south pole; note (b) and (c) are at an earlier time than (a). The `CME legs' and EUV front show a clear spatial separation. Maximum and minimum brightness temperatures in these NRH images are between $5\times10^6$\,K at 445\,MHz and $3\times10^7$\,K at 150\,MHz. The radio sources observed towards the east solar limb are unrelated to the event in question.}
    \label{fig:cme_legs}
    \end{center}
\end{figure}

\subsubsection{Flare Continuum B}

During period 4 in the Orf\'{e}es dynamic spectrum we observe `Flare Continuum B' in Figure~\ref{fig:flux_comparison}(a). It begins at $\sim$500\,MHz and drifts to $\sim$350\,MHz over $\sim$3 minutes. During this time at 408, 432 and 445\,MHz, the previous AR and LT sources have diminished and a new radio source is located inside the north-most flux rope loop (at least in projection on the plane of the sky), shown in Figure~\ref{fig:looptop}(a) -- the loop is highlighted by blue-white crosses because it is obscured by the contours. We label this as `Loop source' or `LS'. The flux density variation of the LS source at 432\,MHz during this time is shown in Figure~\ref{fig:flux_comparison}(e) as the dashed red line, showing it reaches the same peak flux values as the previous AR source at this frequency. 

At frequencies of 327\,MHz and below, there is a source located in the vicinity of the flare site, showing that energised electrons are still being produced close to the flare site. Ongoing flaring activity is also evidenced by the localised area of strong saturation in EUV images and by the presence of FERMI 11--26.9\,keV flux throughout this period in Figure~\ref{fig:flux_comparison} (there is also imaged RHESSI sources at underlying flare ribbons during this time, indicating ongoing flaring activity \citep{joshi2015}). 

Comparing Figure~\ref{fig:looptop}(a) and (b), the 408--445\,MHz show a slight shift westwards over their lifetime, in the same direction of motion as the north-west loop of the rope. The LS source shift is easily observed between 12:54:47\,UT and 12:56:23\,UT in movie\_3.mpg, where just the 432\,MHz contours are shown for simplicity.
We track this slight motion and find position shift by $\sim$40\,Mm over a lifetime of $\sim$100\,seconds, giving a velocity of $\sim$400\,km\,s$^{-1}$ (similar in speed to the LT source discussed above and the speed of the flux rope north-west loop - discussed further in Section~\ref{section:flux_rope_kins}).
The fact that sources from 408-445\,MHz are at the same spatial location and co-propagating with the erupting structure is suggestive that the electron population accelerated at this time starts to fill the internal parts of this structure, quite similar to the observations of \citet{huang2011}. 

After 12:57\,UT all sources then shift back to a location close to the active region, as shown in Figure~\ref{fig:looptop}(c).

Throughout period 3 and 4 the sites of radio emission (energised electrons) are located both close to the flaring active region and moving with the erupting north-west loop of the flux rope (the LT source above the loop followed by the LS source in the loop). This is reminiscent of the simultaneous stationary and moving sources reported in \citet{pick2005}, for example. We discuss our results in the context of such previous observations in Section~\ref{discussion:radio_absorption}.

\subsection{Energetic electrons in CME bubble and legs}

During `period 5', starting at 12:58\,UT in the dynamic spectrum, the radio bursts in Orf\'{e}es become weaker. After this time it is no longer possible to image or define the flux rope loops with AIA. 
Figure 8 (and movie\_4.mpg) shows images of the radio sources using all eight NRH frequencies.
These images were made by applying the three-colour imaging technique to the NRH radio images. We assign one NRH frequency to one colour in an RGB image, allowing us to image three NRH frequencies simultaneously. In this way, spatially coherent regions of intensity are imaged as white, while an excess in intensity of any one (or two) of the NRH frequencies is imaged as a primary colour (or combination of primaries), much the same as the EUV imaging three color technique. 

Figure~\ref{fig:nrh_aia_3col} (a)-(c) show the image triplets at the same time of 12:57:58\,UT -- note this is the same time as Figure~\ref{fig:looptop}(c). Across all frequencies, the sources are now clustered near the flaring active region location, but at a location slightly west of the previous AR source above the flare site. 
Figure~\ref{fig:nrh_aia_3col}(c) shows that at 150, 173, 228\,MHz the radio sources are contained by a roughly circular region in the plane of sky which we call the `radio bubble'. {\color{black}This is probably due to radio emission of energetic electrons localized} in an erupting volume which contains the flux rope. However, because the flux rope is no longer visible in AIA and has yet to emerge in white-light as a CME, we cannot say specifically in which part of this volume the flux rope is contained.
Figure~\ref{fig:nrh_aia_3col}(d) shows that by 12:59:27\,UT the radio emission has spread out to a concave feature which we call the `radio arc', possibly outlining the lower sections of the erupting volume. The three colour AIA image in panel (e) is at the same time as panel (d); it shows a {\color{black}outwardly} propagating EUV front located at the edge of the AIA field of view along with 150\,MHz NRH contour. {\color{black}The radio and EUV feature together show a roughly circular structure; they are most likely related to the boundaries of the erupting volume.}


After $\sim$13:00\,UT the radio sources observed by NRH begin to assemble into two segments, with increasing frequencies towards to the solar surface, see movie\_2.mpg. Figure~\ref{fig:cme_legs}(a) shows the relationship between these two segments (labelled `S1' and `S2') and the CME. The radio contours are overlaid on a 174\,\AA~EUV image from the Sun Watcher using Active Pixel \citep[SWAP;][]{berghmans2006} instrument on board the Project for Onboard Astronomy 2 \citep[PROBA2;][]{santan2013} spacecraft and Large Angle Spectroscopic Coronagraph \citep[LASCO;][]{bru95} C2 instrument on board the Solar and Heliospheric Observatory \citep[SOHO;][]{dom95}. {\color{black}These radio sources converge towards the solar surface and the two segments are arranged in a south-east to north-west orientation with respect to one another; interestingly, this is the same orientation as the original flux rope before eruption, as seen in Figure~\ref{fig:aia_dt_maps} inset}. In order to test if these two segments of radio emission belong to the CME, we compare their positions to the CME front.  However, the closest available LASCO image containing the CME is $\sim$14 minutes after the final NRH image showing the segments. To make a more accurate comparison of the front and radio sources, we determine where the front position was at the time of the radio image. 
This is done by back extrapolating this front along a radial vector towards the parent active region using a speed of 1100\,km\,s$^{-1}$ and travel time of 14 minutes (speed of the CME calculated in Section~\ref{section:cme_speed}). Figure~\ref{fig:cme_legs}(a) shows the positions of the CME front chosen from point-and-click (green crosses), with the back extrapolated front shown as the white-line.
The back-extrapolated front shows a good spatial relationship with the two segments of radio emission, providing evidence that these radio sources delineate the lower sections of the CME. Hence, these two lower sections possibly trace the magnetic roots of the flux rope associated with the CME. We call these magnetic roots the `CME legs', similar to those observed by \citet{huang2011} and \citet{maia1999}.

We also find an interesting relationship between a laterally propagating EUV front in Figure~\ref{fig:cme_legs}(b) and (c) and the position of the CME legs (we call this EUV front `lateral' to distinguish it from the radial front described above). This front first appears along the south limb at at a position angle of $\sim$280$^{\circ}$ at 13:00\,UT. It initially has a speed of $\sim$750\,km\,s$^{-1}$ and decelerates to $\sim$200\,km\,s$^{-1}$ when it then dissipates at $\sim$270$^{\circ}$ in positions angle. It is interesting that the front is situated $\sim$300\,Mm eastwards of the CME legs, which show no discernible motion. It appears the front and legs are two separate physical entities -- we discuss this further in Section~\ref{discussion:cme_legs}. After $\sim$13:15\,UT the radio sources diminish in intensity and lose this roughly two-column structure, marking the end of the radio activity in the metric domain.

\begin{center}
\begin{table*}[!t]
\centering   
\caption{Event kinematics \label{tbl-2}}
    \begin{tabular}{lccccccccc}
        \hline
        	Feature		& Flux Rope 	&  LT source$^\dagger$ 	& Flare Cont.\,B$^*$	& LS source$^\dagger$ 	&  Type II$^*$  		& Outward EUV front 	& CME 		   	\\
	  \hline
	   	T$_{0}$ (UT) &  12:26	&  12:48:27		 &12:53:58 	& 12:55:30	& 12:59:16 	 & 12:56:23	 &  $>$13:25:50 \\
	   	Duration (min) &   29.3	&  	7.6			& 2.4 & 2.0	&  7.8	 	 & 5.6	&  -- \\	
	  
   		 	Speed (km\,s$^{-1}$)&   242$^{+223}_{-100}$	&  395$\pm$80	&  397$^{+71}_{-63}$ &   405$\pm$85	 &  1303$^{+198}_{-140}$	 & 817$\pm$180	& 1140$\pm$380  \\
	  \hline
	    \end{tabular}
	    \tablecomments{Compilation of eruptive kinematics in this event. $T_0$ is the start time of the feature. In the case of the flux rope the quoted speed is the final speed averaged over all d-t traces for the hot and cooler AIA channels. The LT and LS source speeds are from radio imaging. Flare Continuum B and type II speed is derived from frequency drift in dynamic spectra. ~$\dagger$Speed from radio imaging. ~$^*$Speed from frequency drift. }
	  
    \label{tab:kins_table}
\end{table*}
\end{center}

\subsection{Observational summary}

In summary of observations, in period 1 we observe a slowly rising flux rope which becomes destabilised at the time of a C-class flare, type III group (electron beams) and jet. In period 2, during the flux rope acceleration phase, we observe a set of type IIIs above the north-west part of the rope lasting for $\sim$5 minutes. In period 3 we then observe Flare Continuum A in the metric/decimetric domain, which is produced by the stationary source close to the active region. A fainter source is also observed at this time to propagate above the flux rope.
{\color{black}In period 4 we observe Flare Continuum B in dynamic spectra and the associated radio sources in NRH images}. The radio sources are observed to have the same spatial location as the erupting rope north-west loop. In period 5 the radio sources then cluster above the active region and also trace a radio bubble and arc at 150--228\,MHz, eventually spreading {\color{black}to allow two radio segments to be observed simultaneously with the CME seen in white-light}. In general, periods 1-5 reveal where the radio sources (energetic electrons) are located at event initiation and peak, and where they eventually reside when the CME has erupted into the corona.



\section{Kinematic Summary}

In this section we compare the kinematics of each eruptive feature discussed in this event and show the close kinematical relationship between the erupting flux rope and radio sources, summarised in Figure~\ref{fig:total_kins}. 

\subsection{Erupting Flux Rope}
  \label{section:flux_rope_kins}

To derive the kinematics of the flux rope in the early phases of the event, we firstly took intensity traces to build distance-time (d-t) maps from both the hot and cooler AIA passbands over various angles on the erupting structure (Figure~\ref{fig:aia_dt_maps}(b) and (c) shows two d-t maps along one of these angles). We traced the leading edge of these d-t maps and then fit these data with a kinematics profile of $s(t)=s_0 + \alpha t^3$, where $s$ is distance, $t$ is time and $\alpha$ is jerk (time derivative of acceleration). Jerk was used because the distance-time maps cannot be fit with a constant acceleration profile. This implies the {\color{black}erupting structure} was either a constant mass subject to a force which grows with time, a constant force accelerating a mass which depletes with time, or a combination of both. \citet{schrijver2008} also found a jerk profile to fit well the early phase evolution of erupting filaments observed in EUV.

In Figure~\ref{fig:total_kins}, the red and blue lines represent the velocity profiles derived from the $s(t)$ fits from the hot and cooler d-t maps, respectively. The solid and dashed lines represent the velocities from different angles, as indicated by the solid and dashed lines on the inset of the figure (d-t traces at other angles in between these lines showed similar kinematic profiles, just four are shown here for clarity). These show that the first motion to take place was an increase in velocity observed in the hot channels from 1\,km\,s$^{-1}$ at $\sim$12:32\,UT to $\sim$10\,km\,s$^{-1}$ 3 minutes later. At this point the metric type III group, jet and C-class flare occur, as indicated by the vertical pink line on the plot. This shows that a slow motion was occurring before these three signatures of energy release. When this energy release occurs, a rapid increase in velocity begins at the south-east part of the rope, as indicated by the solid lines. The velocity profiles at all other angles over the flux rope show this effect i.e., the acceleration of the rope in every direction begins after the C-class flare just after 12:35\,UT. While the C-class flare, jet and type III group were not the initial cause of motion, they were each linked to the start of the eruption acceleration phase.

Both hot and cool loops accelerate rapidly and reach an average speed over all directions of 242\,km\,s$^{-1}$ after $\sim$12:50\,UT. The top speed of 465\,km\,s$^{-1}$ is calculated in the cooler channels along the solid trace in the inset of Figure~\ref{fig:total_kins}.

\subsection{Kinematics of radio sources and burst drivers}
  \label{section:radio_burst_kins}
  
\begin{figure}[!t]
    \begin{center}
    \includegraphics[scale=0.24, trim=2cm 0.5cm 2cm 1cm]{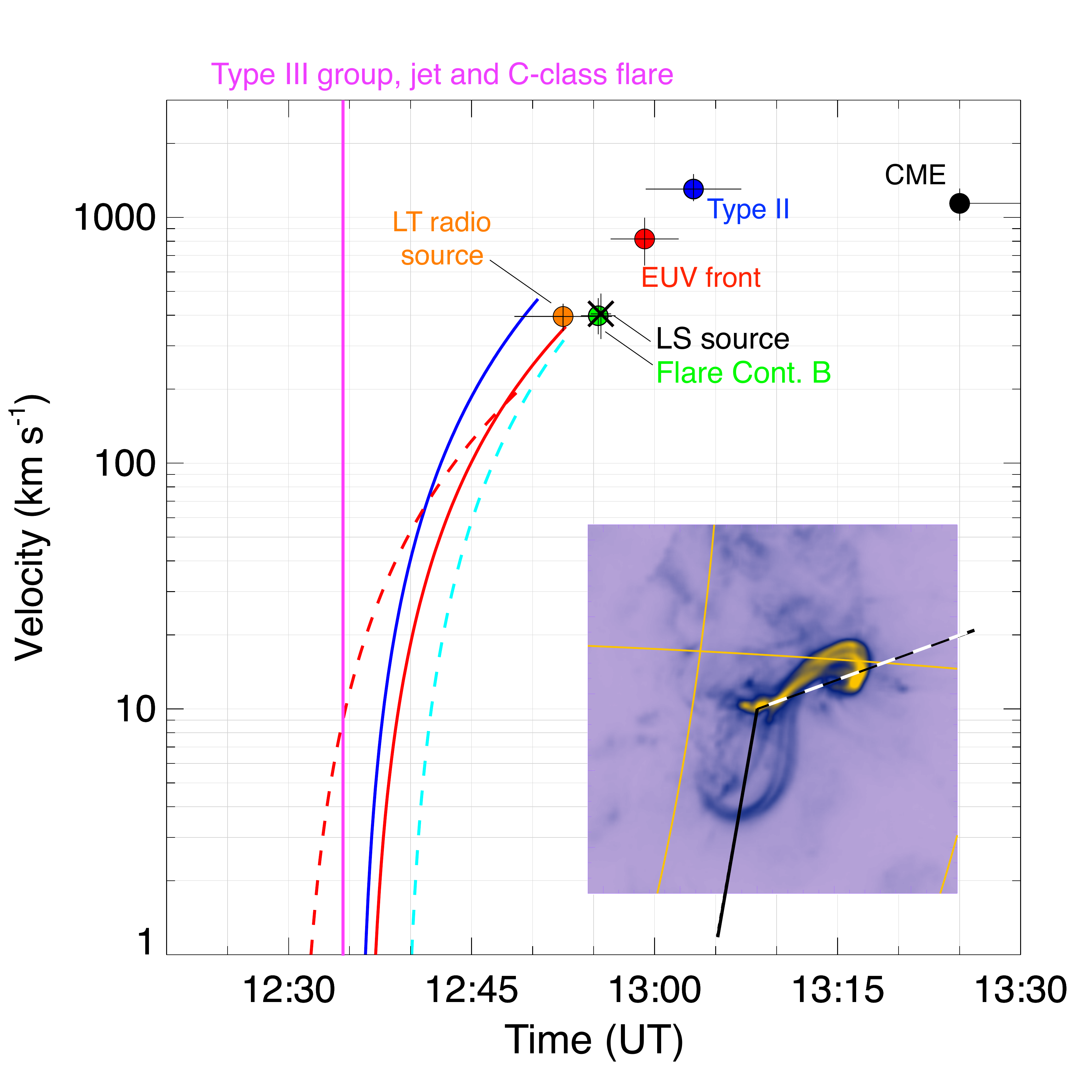}
    \caption{Velocity versus time for eruptive features in the event. The initial slow rise begins in the AIA hot channels to the north-west of the rope, indicated by the dashed red line (d-t trace indicated by the dashed line on the inset). At 12:34:30\,UT the metric type III group, jet and C-class flare occur, shown by the solid vertical pink line. After this is the rapid acceleration on all areas of the flux rope, first beginning with the south-west loop, traced by the solid lines. Blue lines are d-t races from the cooler AIA maps. The LT and LS radio source velocities are derived from images, while the Flare Continuum B and type II velocities are derived from frequency drift in dynamic spectra.}
    \label{fig:total_kins}
    \end{center}
\end{figure}

At $\sim$12:50\,UT we observe the LT radio source above the north-west loop of the rope, with a speed of 395$\pm$80\,km\,s$^{-1}$, calculated from its motion in radio images as shown in Figure~\ref{fig:source_motion}. Similarly the LS source was observed at 12:55\,UT (Figure~\ref{fig:looptop}) at the same location as the north-west loop and had a speed of 405$\pm$85\,km\,s$^{-1}$. The uncertainties quoted for these two speeds are derived from positional uncertainty of 50$''$ for the radio sources (full width half max of the sources), propagated to an uncertainty in velocity. From Figure~\ref{fig:total_kins}, the speed of the flux rope north-west loop (dashed lines) towards the end of the initial jerk stage at $\sim$12:50\,UT is similar to the speeds of the LS and LT radio sources. Hence it is possible this section of the flux rope was the driver of these radio sources.

To derive the speed of the radio bursts in dynamic spectra, we firstly obtain frequency-time points by point-and-click on the radio bursts. These frequency-time points were then used to obtain density-time points assuming harmonic plasma emission and using $f_{plasma}=8980\sqrt{n_e}$, where $f_{plasma}$ is the plasma frequency and $n_e$ is the electron density. We then obtain distance-time points (and speed from a linear fit) from each of the commonly used density models \citep{newkirk1961, saito1977, leblanc1998, mann1999}. 
It is often necessary to multiply the density models by a constant value, e.g. `2-fold Saito', in order to make them appropriate for describing active regions. To search for appropriate model multiplication factors we choose the fold which places the highest frequency in our radio images (445\,MHz) at a height of {\color{black}140\,Mm} i.e., the estimated height at which we observe the NRH 445 MHz above the NW Loop {\color{black}(see Appendix)}. This essentially normalises the models such that they all give the same height for a given initial frequency, resulting in folds of 11.5$\times$Saito, 3.8$\times$Newirk, 21$\times$Leblanc and 18$\times$Mann. These models are then used to derive a set of speeds for a radio burst. From this set of speeds we then compute the mean as the speed of the burst driver and take the minimum and maximum speeds to define the y-error bars in Figure~\ref{fig:total_kins} and Table~\ref{tab:kins_table}. The x-error bars simply represent how long the burst lasted in time. 

Figure~\ref{fig:total_kins} and Table~\ref{tab:kins_table} shows that the Flare Continuum B speed of 397$^{+71}_{-63}$\,km\,s$^{-1}$(from frequency drift) is comparable to the LS source speed 405$\pm85$\,km\,s$^{-1}$(from radio images). 
Furthermore, Flare Continuum B, the LS source, LT source and the north-west flux rope loop (blue dashed line) all have comparable speeds. 
It is likely that this section of the rope interacted with some structure in the surrounding environment and accelerated electrons, causing these radio sources at this location. It is interesting that this location is also close to the origin of the type IIIs observed during period 2. Hence, this region of the corona above the north-west loop had conditions particularly favorable to the production of energetic electrons (resulting in numerous radio sources) as the flux rope erupted. As we will discuss in Section~\ref{discussion:fan_spine}, this location is the predicted site of a magnetic null point in a fan spine structure, making it a likely site of magnetic reconnection \citep{joshi2015}.

\subsection{{\color{black}Outwardly} propagating EUV front and CME}
 \label{section:cme_speed}

Using AIA we tracked the {\color{black}outwardly} propagating EUV front positions off the south-west limb, see Figure~\ref{fig:nrh_aia_3col}(e). Using these positions and an uncertainty of 20\,Mm (approximate uncertainty in identifying the bright feature center in the images), we derived a speed of 817$\pm180$\,km\,s$^{-1}$ for the EUV front. This speed is smaller but comparable to the type II speed of 1303$^{+198}_{-140}$ (derived from frequency drift). Also, we observe the EUV front to be at a radial distance of $\sim$1.2\,R$_{\odot}$ at 12:59\,UT. At this time the type II starts at 150\,MHz, giving the same height of 1.2\,R$_{\odot}$ using the density models quoted above.
{The similar speeds and heights imply a possible kinematical relationship between the EUV and the type II - we discuss this further in Section~\ref{discussion:radio_absorption}.}

Finally we analysed the progression of the CME through the LASCO C2 and C3 fields of view. We computed the position of the front outward from the origin active region up to $\sim$15\,R$_{\odot}$, with error on this position being the extent of the front in each image (between 0.1-0.25\,R$_{\odot}$). Again a simple kinematical fit to the data showed a velocity of 1140$\pm380$\,km\,s$^{-1}$. All features together show an acceleration over time up to the CME measurement in Figure~\ref{fig:total_kins}. However, the CME shows no discernible acceleration during its propagation through the C2 and C3 fields of view.

\begin{figure}[!ht]
    \begin{center}
    \includegraphics[scale=0.25, trim=1cm 4.2cm 2cm 0.7cm]{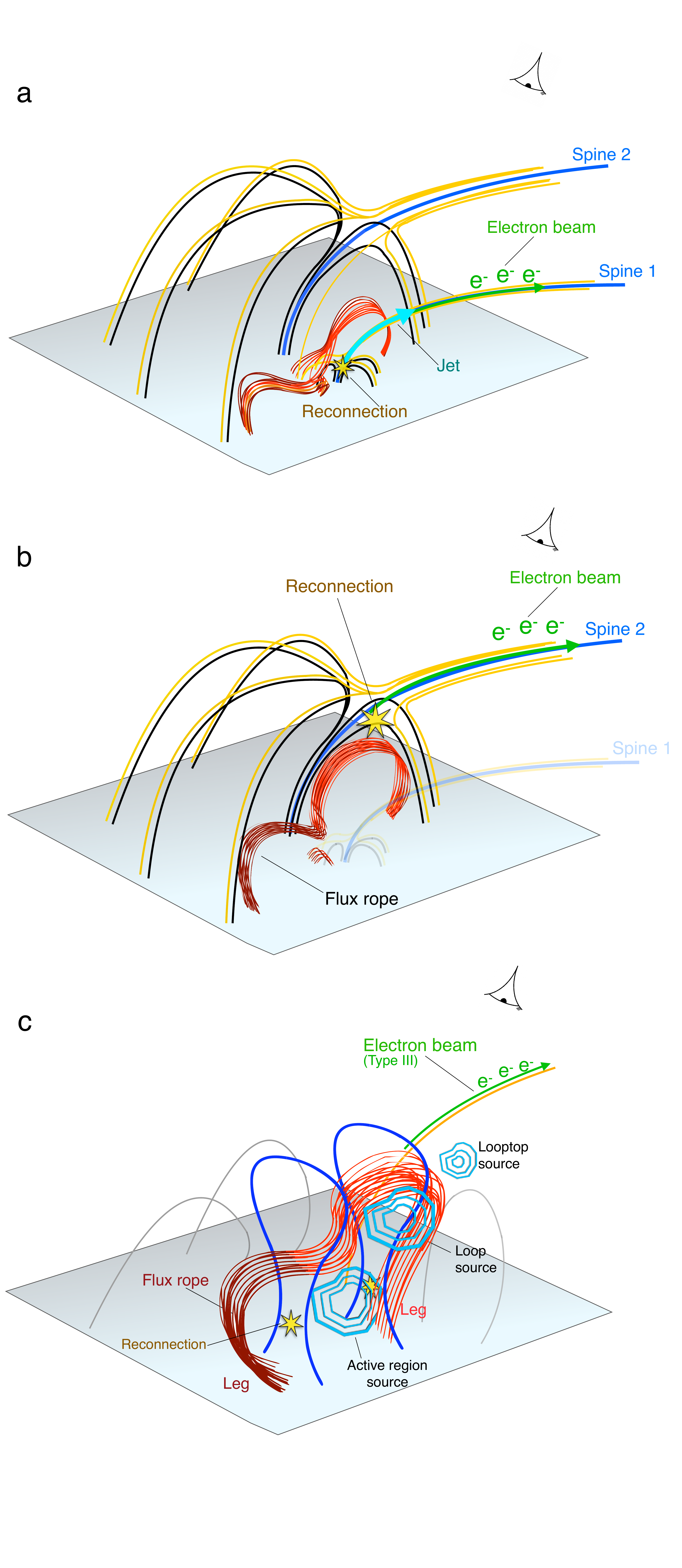}
    \caption{Schematic of event evolution, borrowing the fan-spine structure configuration as outlined by \citet{joshi2015} for this event. (a) Event initiation showing a tether-cutting or flux cancellation-style reconnection close to rope center. The jet and electrons causing the type III group during period 1 may propagate along `Spine 1', which lies close to the center of the flux rope. (b) As the eruption proceeds, one side of the flux rope encounters a null point in the large fan, driving electron acceleration on `Spine 2' and hence producing type IIIs (as observed in period 2). (c) Hypothesised locations of the active region (AR), Looptop (LT) and Loop (LS) sources observed in periods 3 and 4. The electron beam (green line), indicates electrons accelerated at the eruption front.}
    \label{fig:event_evolution}
    \end{center}
\end{figure}

\section{Discussion}

In this section we compare the radio observations of the sites of electron acceleration to what is expected by flux rope eruption models and the previous observations of this event by \cite{joshi2015}. We also discuss the calculation of electron beam energies from the drift rate of the radio bursts during period 1 and 2 of the event. We then discuss the regions amongst the flux rope where the electrons reside towards the end of metric radio activity.

\subsection{Expected sites of electron acceleration during flux rope eruption: Initiation}
	\label{discussion:fan_spine}

During the initiation of the event we have shown that electron acceleration (type III radio sources) was very closely associated with a plasma jet, implying both were caused by the same energy release process. Although this is similar to previous observations of type IIIs and jets \citep{aurass1994, kundu1995, kundu2001, innes2011, chen2013a}, to our knowledge, this is the first such case where the flare, jet and {\color{black}electron acceleration} were so closely related to a flux rope eruption. Whether or not this energy release triggered the eruption or was a result of the eruption is beyond the scope of our analysis. Nonetheless, the site of energy release enables us to analyse the initiation of this event in the context flux rope models.

As shown in Figure~\ref{fig:fluxrope_typeIII}(a) and (b), during event imitation the flare, jet and electron acceleration took place at flux rope center. At this time, the flux rope speed was no more than $10$\,km\,s$^{-1}$ and the jet itself had a relatively small velocity of 120\,km\,s$^{-1}$ \citep{joshi2015}). Given that the Alfv\'{e}n speed at small altitudes in the corona is expected to be $>$1000\,km\,s$^{-1}$ \citep{mann2003}, a shock acceleration mechanism is unlikely, making magnetic reconnection the most probable acceleration mechanism for the electrons. 
Reconnection occurring close to rope center is in support of a tether cutting or flux cancellation scenario \citep{moore2001, vanboo1989}, evidence for which was also provided in detail for this event by \citet{cheng2015} and \citet{joshi2015}. 

A curious feature of this event is that the electron beam escaped from the center of the rope along with a jet. In the classic-tether cutting or flux cancelation scenarios, it is unusual to have an open field line in close vicinity to the rope center. However, from the analysis of \citet{joshi2015}, there is strong evidence that this flux rope was situated close to a fan-spine structure. Figure~\ref{fig:event_evolution} here is a development of such a concept, originally outlined in Figure 13 in \citet{joshi2015}. While their Figure 13 showed a flux rope inside a single fan-spine, we show here the flux rope to be on the boundary of one large fan-spine with a smaller fan-spine located close to the flux rope centre. 
Such a two fan-spine configuration is consistent with the Helioseismic and Magnetic Imager \citep[HMI;][]{schou2012} magnetograms observed before this event (\emph{E. Pariat, 2016, Private Communication}).
Figure~\ref{fig:event_evolution}(a) shows reconnection at the null point of the smaller fan-spine structure occurring close to the flux rope center. The launch of a jet and electron beam along `Spine 1' would explain the presence of the jet and metric type III group during period 1 of our observations. 

{\color{black}It is also possible to estimate the energy of the electrons in the beam launched along with the jet (along Spine 1). We firstly take frequency-time points from the type III group leading-edge in the dynamic spectrum (fastest electrons in the beam which are detectable by the radio burst). We then assume that this is harmonic plasma emission and convert frequency to density via $f_{plasma} \sim 8980\sqrt{n_{e}}$ Hz, with number density $n_{e}$ in cm$^{-3}$. To convert the derived densities to heights (and hence speed and energy) we formulate an appropriate density model using the observed heights of the type III radio sources in Figure~\ref{fig:fluxrope_typeIII}(b), see Appendix.
This results in the use of 1.4$\times$St.\,Hilaire density model. As described in the Appendix, the type III group is actually a type IIId and type IIIn pair observed in Orf\'{e}es and NDA, respectively. Using the St.\,Hilaire model, the drift of the type IIId gives electron energies of {\color{black}$\gtrsim$75\,keV} (the electrons launched with the jet), while the drift of the type IIIn results in $\gtrsim$52\,keV (the electrons escaping to interplanetary distances). However, as described in the Appendix, it is not clear if these two electron populations are one and the same or two independent populations. Hence, the electrons launched with the jet along Spine 1 and those reaching interplanetary distances may be only indirectly related. }

%
%

\subsection{Expected sites of electron acceleration during flux rope eruption: acceleration}

During `period 2', we reported a highly structured set of radio sources which was maintained above the rope north-west loop for over 5 minutes. This clearly indicated a site of electron energization above the rope as it erupts. Figure~\ref{fig:event_evolution}(b) shows a configuration which may explain the observations. Following \citet{joshi2015}, the eruption of the flux rope may have induced reconnection at the null point in the large fan-spine structure, allowing accelerated electrons to escape along `Spine 2' to produce type IIIs\footnote{The estimated energy of the electrons producing these type IIIs is $\sim$5\,keV, found using the drift rate in NDA and the same method as described for the metric type III group.}. In our observations, these type IIIs begin at {\color{black}$\sim$12:42\,UT}. This is exactly the time at which circular flare ribbons (of the proposed fan-spine structure) begin to brighten in EUV images as reported in \citet{joshi2015}. The observations of radio sources above the rope and the brightening of large EUV ribbons lower in the corona is therefore consistent with Figure~\ref{fig:event_evolution}(b) e.g., the electrons propagating on Spine 2 cause the radio emission, while those propagating downwards along the fan produce the EUV ribbons when they reach the low corona/chromosphere. The speed of the erupting rope at the time of this activity was estimated from the distance time maps to be $\sim$100\,km\,s$^{-1}$, see Figure~\ref{fig:total_kins}. As mentioned above, it is unlikely for this speed to be in excess of the Alfv\'{e}n speed low in the corona (on the order of $10^3$\,km\,s$^{-1}$ \citep{mann2003}, {\color{black}ruling out the presence of a shock at this time}. Hence, the most likely electron acceleration mechanism is reconnection above the rope at the null point in the fan-spine structure. {\color{black} A similar interpretation of the interaction of an erupting sigmoid with overlying loops to produce radio pulsations was proposed in \cite{aurass1999}. }  

\subsection{Expected sites of electron acceleration during flux rope eruption: Flare peak}
	\label{discussion:radio_absorption}

We have shown in `period 3' that the stationary radio source close to the active region (AR source) was predominantly responsible for Flare Continuum A seen in the spectrogram. Despite being stationary, an apparent drift of this burst was produced in the dynamic spectrum. {\color{black}This may be due to a number of reasons e.g., the drift in the spectrum may be from the source losing density with time and hence emitting at lower plasma frequencies over time. It could also be caused by the source moving directly toward the observer, resulting in little movement in the plane-of-sky. Despite these possibilities, it remains unclear as to why there was a drift of Flare Continuum A in the dynamic spectrum.}


At the same time as the AR source, we observed an LT source above the northwest section of the rope and moving with a velocity of 395\,km\,s$^{-1}$. The similarity of this to the flux rope velocity lead us to suggest that the flux rope was the driver of this source as it erupted. 
The postulated position of the AR and LT sources with respect to the erupting structure is shown in Figure~\ref{fig:event_evolution}(c) (this is similar to the CSHKP model in three dimensions, with reconnection in a current sheet below the main axis of the flux rope). While the active region source may be located in a reconnection site below the flux rope, the looptop source is driven by the growing flux rope body itself. A similar interpretation of a moving radio source alongside a stationary one was given in \citet{pick2005} i.e., one source close to the post flare loops and another moving with the main body of the flux rope. This is usually observed in imaging as a stationary flare continuum and moving type IV burst \citep{robinson1975, pick1986}. However, in this case the moving radio source (Looptop source) was too weak to be prominent in the dynamic spectrum in this event, so no type IV associated with the LT source was observed in the dynamic spectrum.

An interesting aspect of the type IIIs observed in Figure~\ref{fig:source_motion_imgs} in NDA during period 2 and 3 is the drift of their start frequencies over time, as indicated by the black arrows of the figure. This unusual drift of type III starting frequencies may arise due to the erupting flux rope accelerating electrons on an outer boundary at increasingly larger heights as it erupts into the corona, as indicated on Figure~\ref{fig:event_evolution}(c). It is interesting that the drift of the type III starting frequencies matches the drift of the following type II. This would imply that the boundary which drives type III emission at increasingly larger heights is the same erupting front which drives the type II. Indeed the drift of the type III starting frequency gives a velocity of 1125$^{+136}_{-106}$\,km\,s$^{-1}$ using the density models described above. Within the uncertainty this is comparable to the {\color{black}outwardly} propagating EUV front speed of 817$\pm180$\,km\,s$^{-1}$. {\color{black} Within the uncertainty, it is therefore possible the propagating front was related to the production of the type IIIs -- although the acceleration mechanism of the energetic electrons causing the type IIIs is unclear.} 
{\color{black} Furthermore, given the following type II speed of 1303$^{+198}_{-140}$\,km\,s$^{-1}$, the EUV front could have also been related to the driver of the type II emission. Indeed a front propagating at over 800\,km\,s$^{-1}$ would be expected to drive a shock at 1.1-1.3\,R$_{\odot}$ in the corona given that Alfv\'{e}n speeds may be as low as 200-400\,km\,s$^{-1}$ at this height \citep{mann2003, zucca2014}.
In our case the slightly faster type II speed could arise if it was caused by a piston driven shock i.e., in a piston driven mechanism, the shock (type II) can be faster than its driver (EUV front), as has been simulated in \citet{pomoell2008}. Although is is possible the EUV front is only indirectly related to the type II driver.}

\subsection{Expected sites of electron acceleration during the development of the CME}
	\label{discussion:cme_legs}

Immediately following the AR and LT sources, we observe a `loop source' from 408--445\,MHz to be located in the north-west loop of the erupting flux rope, see Figure~\ref{fig:looptop}(a). The postulated position of this source relative to flux rope is shown in the schematic of Figure~\ref{fig:event_evolution}(c). This implies the electrons accelerated during the eruption are injected inside the erupting flux rope itself. This idea is corroborated by the presence at a later time of the `radio bubble' in Figure~\ref{fig:nrh_aia_3col}, e.g., energetic electrons are now contained by the erupting structure resulting in this circular volume of radio emission. The injection of electrons from a flare site into the {\color{black}erupting structure} has been hypothesized in previous studies to explain the presence of radio emission from within part of the CME itself \citep{bastian2001, maia2007, demoulin2012}. In our study, Figure~\ref{fig:looptop}(c) seems to be the early stages of such a process i.e., we observe radio sources to be co-located and co-propagating with the erupting loops of the flux rope when it is still quite low in the corona. These electrons are then imaged to cover a large volume of the eruption in the form of the radio bubble and arc in Figure~\ref{fig:nrh_aia_3col}. 

Following this, radio sources form two segments which converge on the active region, see Figure~\ref{fig:cme_legs}. We postulate that these radio sources are from energetic electrons which are fixed to the magnetic field of the flux rope legs as it erupts into the corona and forms a CME. The hypothesised leg positions are indicated on Figure~\ref{fig:event_evolution}(c). There have previously been some evidence that CME legs exist in radio observations \citep{maia1999, huang2011, dressing2016}, however these cases were either low in the corona or constrained to just one frequency of observation. Here our observations reveal the radio sources to extend into the corona and show close spatial relationship with the back-extrapolated CME front in white-light. 

Finally, there is much debate on the relationship between EUV waves and CMEs \citep{gallagher2011}. A prevailing hypothesis is that these features are fast mode magnetohydrodynamic waves propagating through the corona \citep{mann1999a, veronig2010}, possibly driven by the CME eruption. However, there is also a `pseudo-wave' interpretation, whereby the erupting CME produces a large-scale restructuring or reconnection of coronal magnetic field \citep{chen2002, attrill2007}. 
Recent investigations have also suggested a hybrid between the wave and pseudo-wave theories \citep{liu2012, downs2012}. However, the difficulty in imaging both the EUV wave and CME simultaneously at low altitudes makes distinguishing these two phenomena problematic and there is still no consensus as to the nature of EUV waves. Figure~\ref{fig:cme_legs}(b) and (c) show that the laterally propagating EUV front in particular may be freely propagating and entirely separate from the CME structure itself. Alternatively, it may also be interpreted as the outer boundary of the CME while the legs represent some internal part of the erupting structure.

\section{Conclusion}

In this study, {\color{black}we have presented analysis of radio sources which indicate the regions of accelerated electrons at each stage of a flare and flux rope eruption}, from its initiation to its propagation into the corona. While the flux rope was observed in AIA, the sites of electron acceleration were identified using multiple radio frequency images of the Nan\c{c}ay Radioheliograph. This combined with high time and frequency resolution dynamic spectral observations from Nan\c{c}ay's new Orf\'{e}es instrument allowed us to identify when and where the electron acceleration took place during the event. Our observations and analysis reveal the following properties of electron acceleration sites during flux rope eruption in this event:

\begin{enumerate}

 \item At the time of flux rope eruption, tether-cutting or flux cancelation-type reconnection takes place at flux rope center resulting in the expulsion of a plasma jet and the acceleration of electron beams to {\color{black}$\gtrsim$75\,keV} which escape into the corona and produce a type IIId radio burst. The escape of these beams requires the null point of a fan-spine structure to be located close to the rope center, with electrons propagating along the spine. At the same time, the presence of a type IIIn burst indicates $\gtrsim$52\,keV electrons propagating to interplanetary distances. However, it is unclear as to the relationship between the type IIId and type IIIn. The two different radio bursts may be from two separate populations of electrons, so there may only be an indirect link between the population of electrons accelerated along with the jet, and those escaping into the heliosphere.
 
 \item As the flux rope erupts, reconnection takes place above the rope, resulting in repeated acceleration of {\color{black}electrons of energies of 5\,keV} for a period of up to 5 minutes. The site of reconnection is likely at a null point in a large fan-spine structure above the flux rope in the corona. The electrons accelerated during this time escape into the corona along the spine of this structure and produce type III radio bursts. This observation is in support of the hypothesised points of reconnection (implying electron acceleration) at the fan-spine null-point outlined in \citet{joshi2015}.
 
 \item During flare peak, the majority of the electron acceleration takes place close to the flaring active region. Simultaneously, reconnection driven above the rope as it erupts into the corona results in a site of further electron acceleration which propagates outwards at the same speed as the rope ($\sim$400\,km\,s$^{-1}$). Following this, electron acceleration continues in the active region and electrons are injected onto the loops of the rope.
 
 \item {\color{black}Electron acceleration} continues close to the flare site. During this time we find evidence of energetic electrons beginning to fill the erupting volume. The growth of this volume results in electrons being contained on the magnetic fields that make up the legs of the CME, allowing the legs to be clearly imaged at multiple frequencies high in the corona. In future, low frequency imaging spectroscopy, such as that now provided by instruments like the Low Frequency Array \citep[LOFAR;][]{vanHaarlem2013}, may reveal just how far CME legs may be imaged in the corona e.g., as low as 30\,MHz.
\end{enumerate}

\acknowledgments{Eoin Carley is supported by ELEVATE: Irish Research Council International Career Development Fellowship -- co-funded by Marie Curie Actions. We would like to thank Sophie Masson and Etienne Pariat for some very useful discussion of this event. We would also like to thank the referee for their very useful and constructive comments. We are grateful to the SDO, SOHO, SWAP, RHESSI, FERMI, and GOES teams for open access to their data. The NRH is funded by the French Ministry of Education and the R\'{e}gion Centre. Orf\'{e}es is part of the FEDOME project, partly funded by the French Ministry of Defense.}

\appendix
\section{Height estimate of radio sources}

 \begin{figure}[!b]
   \begin{center}
   \includegraphics[scale=0.3,  trim=0cm 4cm 0cm 0.0cm]{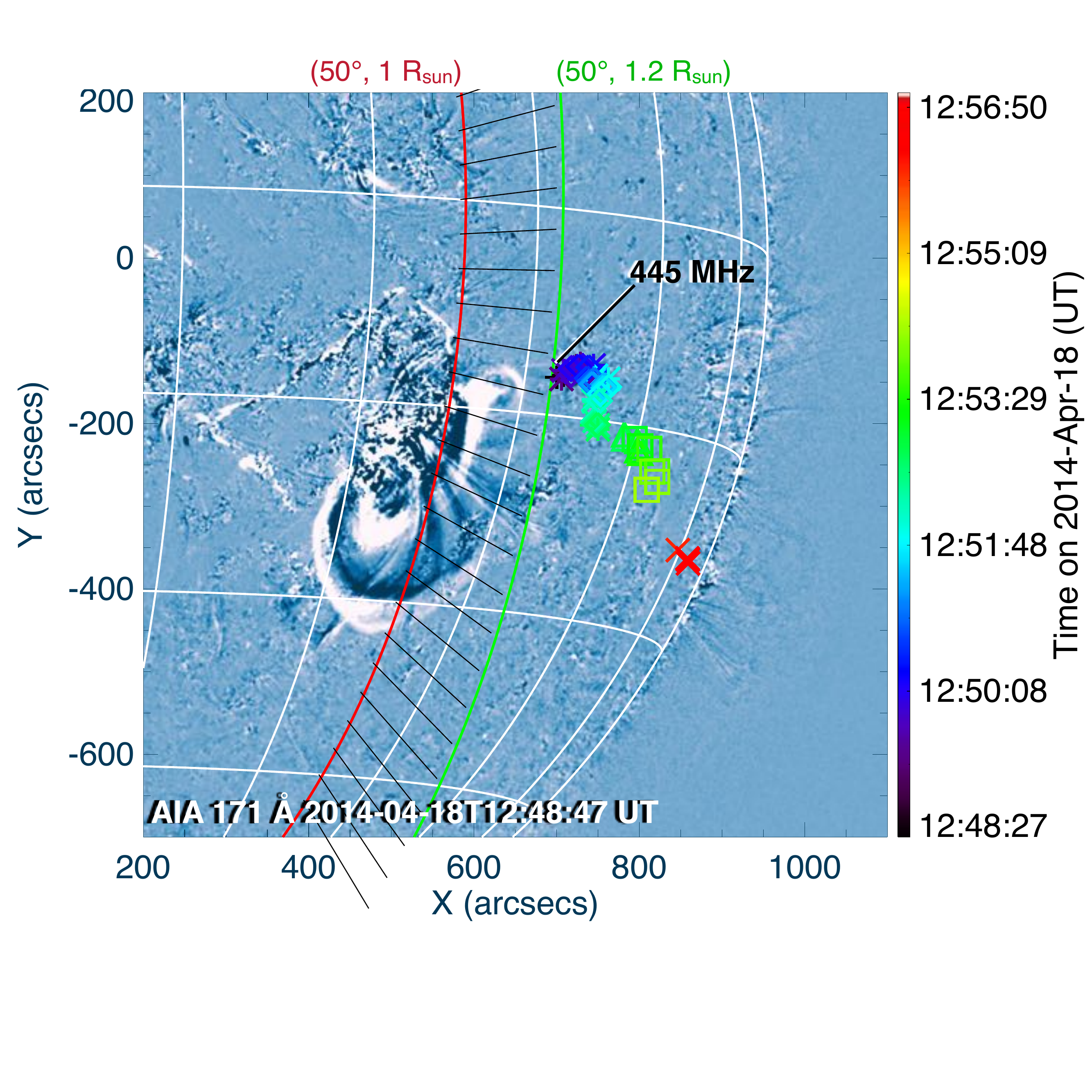}
    \caption{{\color{black}AIA 171\,\AA~running difference image of the erupting flux rope. The starting position of the 445\,MHz source is indicated. We assume the NW loop and the 445\,MHz source are on the same plane e.g., one which makes an angle of 50$^{\circ}$ with the POS. This then allows us to determine the height of the 445\,MHz to be 1.2\,R$_{\odot}$}.}
    \label{fig:source_motion_appendix}
    \end{center}
\end{figure}

{\color{black} Since we have the possibility to observe the actual heights of radio sources in the corona from imaging, the density models which we use can be altered such that they produce results that are consistent with images. This procedure essentially `normalises' the models such that they each give the same height for a particular density (frequency), with this height matching the observed source height (at the same frequency) in radio images. This was done both for deriving a speed of Flare Continuum B and the type III group in our observations:

\subsection{1. Normalization heights for Flare Continuum B}
Figure~\ref{fig:source_motion_appendix} shows how a normalization height was estimated for 445\,MHz during Flare Continuum B.
The figure shows the erupting flux rope observed using an AIA\,171\,\AA~running difference image, overlaid with the LT source positions in the corona as imaged by all NRH frequencies over time (the same as Figure~\ref{fig:source_motion}). The initial position of the 445 MHz source is indicated. If we assume the NW Loop footpoint and 445 MHz source position are in the same plane (i.e., the radio source lies radially above the NW loop) we can estimate the radio source height. The pre-eruptive position of the flux-rope NW loop footpoint is $\sim$50$^{\circ}$ from the plane of sky (POS). The red line shows a trace at this POS angle along the solar surface (heliocentric distance of 1\,R$_{\odot}$). 
The green line shows a longitude trace at the same POS angle at a heliocentric distance of 1.2\,R$_{\odot}$. The black lines then define the plane between the red and green lines -- this plane is perpendicular to the solar surface. This places the height of the 445\,MHz source at 140 Mm or 1.2\,R$_{\odot}$ (the height of the green curve). 
Assuming this is harmonic emission, such a height for this frequency requires density models of the form 11.5$\times$Saito, 3.8$\times$Newirk, 21$\times$Leblanc and 18$\times$Mann. The use of these density models give an average velocity of 397\,km\,s$^{-1}$ for the Flare Continuum B frequency drift. This matches the velocity of 405\,km\,s$^{-1}$ for the associated radio sources in the images, showing that the quoted density models provide reasonable speeds for this drifting radio burst.

\subsection{2. Normalization heights for Type III group}
The same procedure was performed for the type III radio sources, see Figure~\ref{fig:typeIII_appendix}. Assuming the 298\,MHz source lies in the same plane as the jet and initial C-class flare (which took place at a longitude of $\sim$52$^{\circ}$ from the POS), the height of this radio source is 1.16\,R$_{\odot}$. In order to convert from drift to beam speed, we chose the St.\,Hilaire density model \citep{sthilaire2013}. This is a solar wind-like density model of the form $n_e(r) = C(h/R_{\odot})^{-2}$, where $h$ is heliocentric distance and C is a normalizing coefficient that was found from a statistical analysis of over over 8000 type III events using  NRH. Because it is a generic solar wind-like model, it is appropriate for describing the density gradient in open-field structures at any distance in the corona and solar wind. A height of 1.16\,R$_{\odot}$ for 298\,MHz harmonic emission requires a 1.4$\times$St-Hilaire model, showing that this model is consistent with the height at which we see the type III source in our observations. 

 \begin{figure}[!t]
   \begin{center}
   \includegraphics[scale=0.22,  trim=0cm 2cm 0cm 0cm]{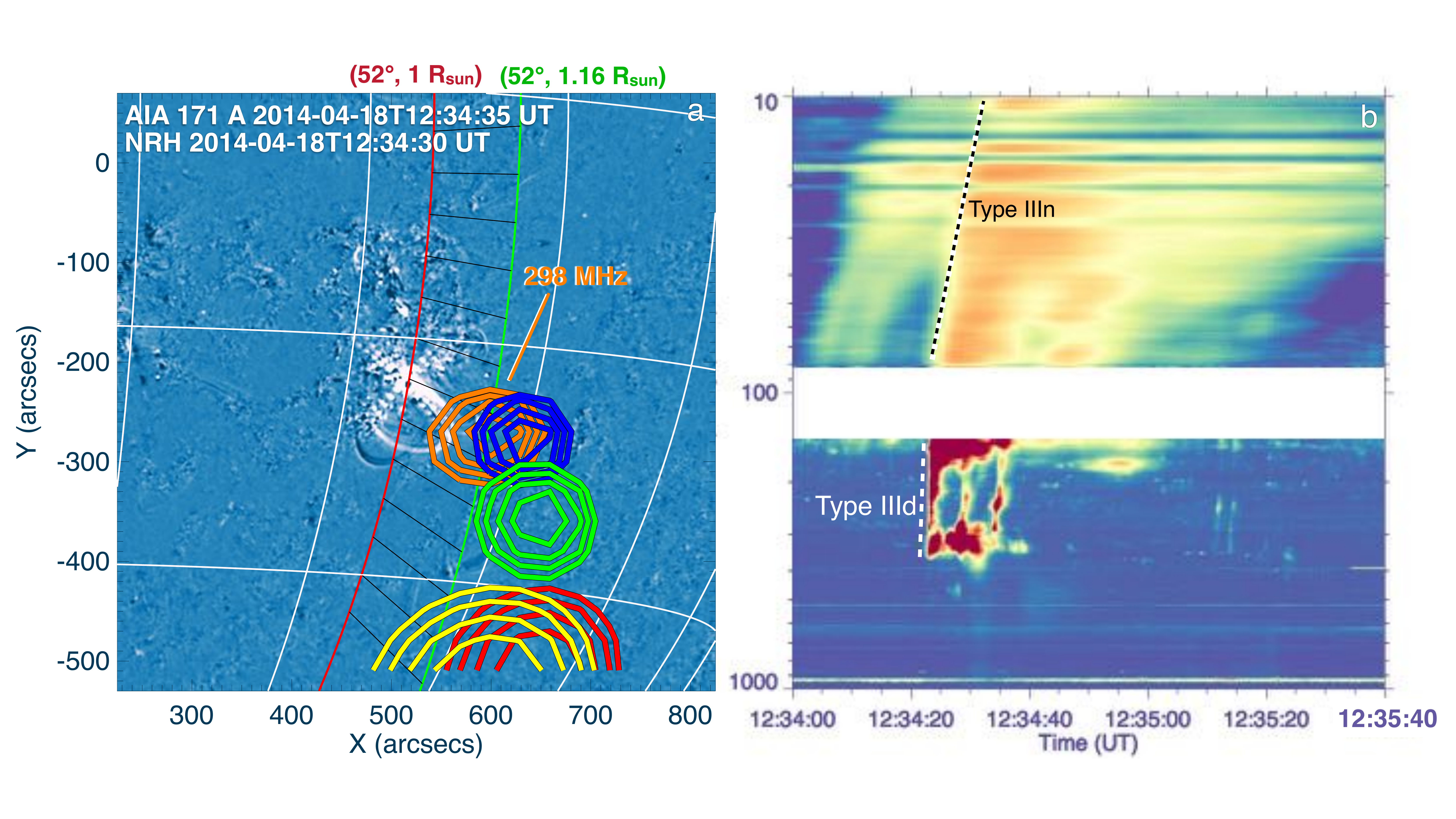}
    \caption{(a) AIA 171\,\AA~running difference image overlaid with the NRH contours of the type III radio burst at 12:34:30 UT. The 298 MHz source is indicated. We assume the flare site, jet and 298 MHz source are on the same plane at 52$^{\circ}$ from the POS e.g., the radio source lies above the flare and jet. This then allows us to determine the height of the 298 MHz source to be 1.16\,R$_{\odot}$, resulting in a normalsied 1.4$\times$St.\,Hilaire density model. This model is then used to derive a speed of the associated type III radio burst in the dynamic spectra. (b) Type III group observed with Orf\'{e}es and NDA, showing that this is a type IIId and type IIIn pair.}
      \label{fig:typeIII_appendix}
    \end{center}
\end{figure}

Now, in Figure~\ref{fig:typeIII_appendix}(b) the type III drift is $\sim$-368\,MHz\,s$^{-1}$ in Orf\'{e}{e}s and $\sim$-13\,MHz\,s$^{-1}$ in NDA. The dramatically different drift rates between high and low frequency means we are observing a combination of type IIIn + type IIId such as outlined in \citet{poquerusse1994} and \citet{klassen2003b}. Firstly, for the type IIId we obtain a speed of $\sim$0.96\,c using the 1.4$\times$St.\,Hilaire model. As expected for type IIId bursts in general, 
the speed is excessive and is due to electron time-of-flight effects combined with a finite light-speed and a propagation path towards the observer. The abnormally high drift (and hence speeds) of the burst can be corrected to find the real speed of the electron beam, if we know the angle of propagation between beam direction and line of sight (LOS). We do not have this angle, however following the analysis of \citet{klassen2003b} (their Section\,5) we may estimate the minimum possible real electron speed $v_{min}$ capable of producing the apparent speed $v_{app}$
\begin{equation}
v_{min}=\frac{c\,v_{app}}{c + v_{app}}
\end{equation}
where $c$ is the speed of light. The minimum real beam speed is then $v_{min}=0.49$\,c (75\,keV) for propagation directly towards the observer. Hence, despite not knowing the LOS angle, the advantage of observing a type IIId is that it allows us to place a minimum on the actual speed of the beam. Furthermore, the beam is unlikely to propagate directly towards the observer, so the speed is likely larger than 0.49\,c.

Given the type IIIn drift  of $\sim$13\,MHz\,s$^{-1}$, the speed derived from this drift is ~0.71c using the normalized St. Hilaire model. Again, the correction factor gives a minimum possible speed of 0.42\,c (52\,keV). This speed is typical of previous speeds derived type IIIn bursts and is faster (but not untypical) than a `normal' type III radio burst. In general, it is unclear as to the cause of the sudden drift-rate change between type IIId and type IIIn bursts. While some theories suggest the two separate bursts are from two different instabilities arising from the same energetic electron population \citet{klassen2003b}, others suggest that the two separate bursts are from two separate populations of electrons \citep{benz1982}. It is therefore not possible to say if the electrons accelerated along with the jet (type IIId) are those that escape to interplanetary distances (type IIIn) -- we can only say they are closely related.

}

\vspace{10mm}


\end{document}